%% file: main.tex
\documentclass[%
aip,%
amsmath,amssymb,floatfix,
reprint,%
onecolumn,%
]{revtex4-1}

\usepackage{type1ec} 

\usepackage{amsmath,amsthm,latexsym,amssymb,amsfonts,epsfig}

\usepackage[english]{babel}
\usepackage[utf8]{inputenc}			
\usepackage{graphicx, texdraw}   
\usepackage{bm}

\usepackage{mathrsfs}

\usepackage{mathtools}  
\usepackage{makecell}   

\DeclarePairedDelimiter{\ceil}{\lceil}{\rceil}

\usepackage[left= 2.2cm, right = 2.2cm, top= 2.cm, bottom = 2.5cm]{geometry}

\usepackage{enumerate}

\usepackage[colorlinks=true,citecolor=OliveGreen, linkcolor=blue]{hyperref}

\usepackage{csquotes} 

\usepackage[usenames, dvipsnames]{color}

\usepackage{amsmath}
\usepackage{amsfonts}
\usepackage{amssymb}
\usepackage{braket}

\usepackage{float}
\floatstyle{plaintop}
\restylefloat{table}

\input{commands}

\begin{document}
\title{Axisymmetric Flow due to a Stokeslet Near a Finite-Sized Elastic Membrane}

\author{Abdallah Daddi-Moussa-Ider}
\email{abdallah.daddi.moussa.ider@uni-duesseldorf.de}
\affiliation
{Institut f\"{u}r Theoretische Physik II: Weiche Materie, Heinrich-Heine-Universit\"{a}t D\"{u}sseldorf, Universit\"{a}tsstra\ss e 1, 40225 D\"{u}sseldorf, Germany}

\author{Badr Kaoui}
\email{badr.kaoui@utc.fr}
\affiliation{
Biomechanics and Bioengineering Laboratory (UMR 7338), CNRS, Universit\'{e} de Technologie de Compi\`{e}gne, 60200 Compi\`{e}gne, France}

\author{Hartmut L\"{o}wen}
\email{hartmut.loewen@uni-duesseldorf.de}
\affiliation
{Institut f\"{u}r Theoretische Physik II: Weiche Materie, Heinrich-Heine-Universit\"{a}t D\"{u}sseldorf, Universit\"{a}tsstra\ss e 1, 40225 D\"{u}sseldorf, Germany}

\date{\today}


\begin{abstract}

	Elastic confinements play an important role in many soft matter systems and affect the transport properties of suspended particles in viscous flow.
	On the basis of low-Reynolds-number hydrodynamics, we present an analytical theory of the axisymmetric flow induced by a point-force singularity (Stokeslet) directed along the symmetry axis of a finite-sized circular elastic membrane endowed with resistance toward stretching, area expansion, and bending.
	The solution for the viscous incompressible flow surrounding the membrane is formulated as a mixed boundary value problem, which is then reduced into a system of dual integral equations on the inner and outer sides of the domain boundary.
	We show that the solution of the elastohydrodynamic problem can conveniently be expressed in terms of a set of inhomogeneous Fredholm integral equations of the second kind with logarithmic kernel.
	Basing on the hydrodynamic flow field, we obtain semi-analytical expressions of the hydrodynamic mobility function for the translational motion perpendicular to a circular membrane.
	The results are valid to leading-order in the ratio of particle radius to the distance separating the particle from the membrane.
	In the quasi-steady limit, we find that the particle mobility near a finite-sized membrane is always larger than that predicted near a no-slip disk of the same size.
	We further show that the bending-related contribution to the hydrodynamic mobility increases monotonically upon decreasing the membrane size, whereas the shear-related contribution displays a minimum value when the particle-membrane distance is equal to the membrane radius.
	Accordingly, the system behavior may be shear or bending dominated, depending on the geometric and elastic properties of the system. 
	Our results may find applications in the field of nanoparticle-based sensing and drug delivery systems near elastic cell membranes.

\end{abstract}
\maketitle

\section{Introduction}

Geometric confinements are commonly encountered in soft matter systems and in particular affect the behavior and transport properties of colloidal suspensions in a viscous medium~\cite{waigh16, graham11, yamamoto08, nakayama05, diamant09, molina13, kusters14}.
Hydrodynamic interactions between nanoparticles and elastic cell membranes play a key role in a large variety of biological and technological applications.
A notable example being targeted drug delivery using nanoparticle carrier systems, which navigate through blood vessels to reach disease sites such as tumors and inflammation areas~\cite{veiseh10, naahidi13, al-obaidi15, liu16}.
During their uptake by living cells via endocytosis~\cite{Doherty_2009, meinel14, AgudoCanalejo_2015}, the behavior of nanoparticles is strongly affected by hydrodynamic interactions with living cells composing of lipid and protein membranes.

In low-Reynolds-number hydrodynamics, the presence of nearby interfaces is known to drastically modify the flow field around immersed objects because of the long-range nature of the fluid-mediated hydrodynamic interactions. 
Over the last few decades, considerable research efforts have been devoted to the theoretical study of the slow (creeping) motion of a small particle moving close to a rigid boundary~\cite{mackay61, gotoh82, cichocki98, lauga05, swan07, franosch09, swan10, felderhof12, decorato15, huang15, rallabandi17obstacle}, a fluid-fluid interface separating two immiscible viscous fluids~\cite{lee79, berdan81, bickel06, bickel07, blawz10theory, blawz10}, or a soft membrane endowed with surface elasticity~\cite{felderhof06, shlomovitz14, daddi16, daddi16b, daddi16c, daddi18coupling}, 
finding that particle diffusion perpendicular to the interface is significantly hindered compared to that along the direction parallel to the interface.

Unlike a solid-liquid or a liquid-liquid interface, an elastic interface stands apart as it  introduces a memory effect in the system that causes a long-lasting anomalous subdiffusive behavior on nearby particles~\cite{daddi16}.
In addition, a rigid object that is dragged tangent to an elastic wall in a viscous fluid experiences a lift force that is directed normal to its direction of motion~\cite{skotheim04, skotheim05, yin05, snoeijer13, wang15, wang17, wang17SM}.
This lift mechanism is a direct consequence of the elastic deformation of the interface, which breaks the time-reversal symmetry of Stokes flows.
Experimentally, particle motion near confining interfaces has been investigated using optical tweezers~\cite{faucheux94, dufresne01, schaffer07}, video microscopy~\cite{dufresne00, eral10, cervantesMartinez11, dettmer14}, or evanescent wave dynamic light scattering~\cite{holmqvist07, michailidou09, lisicki12, rogers12, michailidou13, lisicki14}.
Meanwhile, the influence of a nearby elastic interface has been investigated using magnetic particle actuation~\cite{irmscher12}, optical traps~\cite{shlomovitz13, boatwright14, junger15, traenkle16}, or rotating coherent scattering microscopy~\cite{junger16}.

In this paper, we calculate theoretically the axisymmetric flow field induced by a Stokeslet situated along the axis of a finite-sized circular membrane possessing resistance toward shear and bending.
The solution of the flow equations is formulated as a mixed boundary value problem, which is then reduced to a set of convergent Fredholm integral equations of the second kind.
We then derive semi-analytical expressions of the frequency-dependent mobility function that relates the velocity of a particle moving near a membrane to the hydrodynamic force exerted on its surface.
More prominently, we show that the system behavior may be dominated by shear or bending, and that depending on the membrane size and the distance separating the particle from the membrane.

The remaining of the paper is organized as follows.
In Sec.~\ref{sec:problemStatement}, we state the elastohydrodynamic problem and introduce a relevant model for the membrane that incorporates both shear and bending deformation modes.
We then formulate in Sec.~\ref{sec:formulationSolutionFlowProblem} the solution of the flow equations and derive the corresponding mixed boundary value problem which we solve in Sec.~\ref{sec:solutionMixedBoundaryValueProblem} for idealized membrane with pure shear or pure bending.
In Sec.~\ref{sec:hydrodynamicMobility}, we provide semi-analytical expressions of the hydrodynamic mobility functions.
Concluding remarks are contained in Sec.~\ref{sec:conclusions}.
Further technical details which are not essential to the understanding of results are relegated to appendices.

\section{Mathematical formulation}
\label{sec:problemStatement}

\begin{figure}
	\centering
	\includegraphics[scale=1.5]{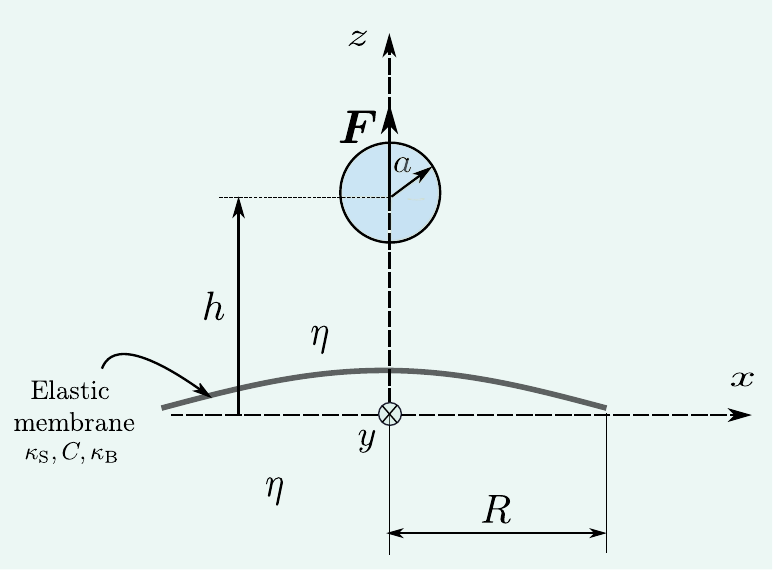}
	\caption{(Color online) Cross section of the system setup. 
	A solid spherical particle of radius~$a$ translating perpendicular to a finite-sized elastic membrane of radius~$R$, under the action of an external force~$\vect{F}$. 
	The surrounding fluid is Newtonian characterized by a constant dynamic viscosity~$\eta$.
	The membrane is composed of a hyperelastic material that possesses resistance toward shear elasticity and bending stiffness.
	}
	\label{systemSetup}
\end{figure}

We consider the axisymmetric motion of a solid spherical particle of radius~$a$, initially located a distance~$h$ above a finite-sized membrane of radius~$R$ that is extended in the~$xy$ plane.
The unit vector~$\ez$ is directed normal to the undeformed membrane.
The particle is moving under the action of an arbitrary time-dependent external force~$\vect{F}(t) = F(t) \, \ez$, as schematically illustrated in Fig.~\ref{systemSetup}.
Here, we investigate the system behavior in the far-field limit, where the particle size is small compared to the particle-membrane distance.
The fluid is Newtonian of constant dynamic viscosity~$\eta$ and the flow is assumed to be incompressible.
The membrane is modeled as a two-dimensional hyperelastic circular disk endowed with resistance toward shear and bending.
The shear elasticity of the membrane is described by the Skalak model~\cite{skalak73}, which
incorporates both the resistance toward shear and area dilation.
The Skalak model is characterized by the elastic shear modulus~$\kS$ and the area expansion modulus~$\kA$, both of which are related by the dimensionless number $C=\kA/\kS$.
The latter is commonly known as the Skalak parameter~\cite{kaoui12, Freund_2014}, and is typically very large for red blood cells, so as to express the area incompressibility of cell membranes.
The resistance of the membrane toward bending is modeled by the well-established Helfrich model, which is commonly used as a relevant model for bilayer lipid vesicles or liposomes.

For a small displacement of the membrane relative to a horizontal plane of reference, the traction jump equations arising from these two deformation modes are expressed in a linearized form by~\cite{daddi16, daddi16b, daddi16c}
\begin{subequations}\label{tractionJumpEqs}
	\begin{align}
	-\frac{\kS}{3} \big( \Delta_{\parallel} u_\beta + (1+2C) \epsilon_{,\beta} \big) &= \bigtriangleup    f_{\beta} \, , \qquad \beta \in \{x,y\} \, , \label{sigma_tangential}  \\
	 \kB   \Delta_{\parallel}^2 u_z &= \bigtriangleup    f_{z}  \quad \text{on} \quad \RS \, , \label{sigma_normal}
	\end{align}
\end{subequations}
where $\vect{u}$ denotes the membrane displacement field and $\bigtriangleup  \vect{f}$ is the traction jump vector across the membrane.
Moreover, $\Delta_\parallel$ is the Laplace-Beltrami operator~\cite{deserno15}, which, for a given function~$g(x,y)$, is defined in Cartesian coordinates as $\Delta_\parallel g := g_{,xx} + g_{,yy}$.
In addition, $\epsilon:= u_{x,x}+u_{y,y}$ is the dilatation function, and $\RS = x \, \ex + y \, \ey$ denotes the position vector of the material points of the membrane in the planar configuration of reference. 
We note that a comma in indices represents a partial spatial derivative with respect to the corresponding variable.

Assuming low-Reynolds-number hydrodynamics~\cite{leal80, happel12, kim13}, the fluid velocity and viscous stress fields, respectively denoted as~$\vect{v}(\R, t)$ and $\vect{\sigma} (\R,t)$, satisfy the stationary Stokes equations
\begin{subequations}\label{StokesGleischungen}
	\begin{align}
		\bNabla \cdot \vect{v} &= 0 \, , \\
		\bNabla \cdot \vect{\sigma} + \vect{f} &= 0 \, , 
	\end{align}
\end{subequations}
wherein~$\vect{f}$ is the force density acting on the surrounding fluid due to the presence of the suspended particle.
The total force~$\vect{F}$ is obtained by integrating over the particle surface.
The viscous stress tensor is $\vect{\sigma} = -p \vect{I} + 2\eta \vect{E}$ where $p$ denotes the pressure field and $\vect{E} = \tfrac{1}{2} \left( \bNabla \vect{v} + \bNabla \vect{v}^{\mathrm{T}} \right)$ is the rate-of-strain tensor.
Moreover, the components of the traction jump vector appearing on the right-hand side of Eqs.~\eqref{tractionJumpEqs}, are related to the fluid stress tensor by
\begin{equation}
	\bigtriangleup f_\beta = \sigma_{z\beta} (z=0^+) - \sigma_{z\beta} (z=0^-) \, , 
	\qquad \beta \in \{x,y,z\} \, .
\end{equation}

It is worth mentioning that we have omitted the unsteady term in the Stokes equations since, in realistic situations, it yields a negligible contribution to the induced flow field near elastic interfaces~\cite{daddi16}.

In the following, we approximate the hydrodynamic force by its first multipole moment such that $\vect{f} (t) = \vect{F} (t) \, \delta(\R-\R_0)$.
Accordingly, the elastohydrodynamic problem reduces to solving the governing equations of fluid motion for a point-force singularity positioned at the center of the particle and located above the planar membrane at position~$\R_0$.

Since the flow is axisymmetric, the problem can more conveniently be solved using cylindrical coordinates.
The stationary Stokes equations governing the fluid motion in the cylindrical coordinate system read~\cite{kim13}
\begin{subequations}\label{StokesGleischungenCylind}
	\begin{align}
		\frac{v_r}{r} + v_{r,r} + v_{z,z} &= 0 \, , \\
		-\frac{p_{,r}}{\eta} + \Delta v_r - \frac{v_r}{r^2} &= 0 \, , \\
		-\frac{p_{,z}}{\eta} + \Delta v_z + F \, \delta (z-h) &= 0 \, ,
	\end{align}
\end{subequations}
where~$\Delta$ is the axisymmetric Laplace operator, which, for a function~$g(r,z)$ expressed in cylindrical coordinates, is given by $\Delta g = g_{,rr} + g_{,r}/r + g_{,zz}$.

\section{Formulation of the solution of the flow problem}
\label{sec:formulationSolutionFlowProblem}

\subsection{Solution form}

Due to the linearity of the Stokes equations, the solution of the flow problem can be written as a linear superposition of the free-space Green's function and a complementary solution that is required to satisfy the boundary conditions prescribed at the elastic membrane.
In an unbounded fluid, the radial and axial velocities due to a Stokeslet are~\cite{kim13}
\begin{equation} \label{stokesletGeschwindigkeitGleischung}
 \vStokcom_r = \frac{F}{8\pi\eta} \frac{r(z-h)}{s^3} \, , \qquad 
 \vStokcom_z = \frac{F}{8\pi\eta} \left( \frac{2}{s}-\frac{r^2}{s^3} \right) \, , 
\end{equation}
where $r=\left(x^2+y^2\right)^{1/2}$ and $s= |\R-\R_0|$ is the distance from the Stokeslet position.
In the far-field limit, the flow velocity field decays as $1/s$.
This implies that hydrodynamic interactions are long ranged and can strongly be altered by geometric confinements. 
The corresponding Stokeslet solution for the pressure field reads
\begin{equation}\label{stokesletDruckGleischung}
	\pStok = \frac{F}{4\pi} \frac{z-h}{s^3} \, .
\end{equation}

In the presence of a confining interface, the solution of the flow problem for the velocity and pressure fields can be presented in the form
\begin{equation}\label{GeneralSolution}
	\vect{v} = \vect{v}^\mathrm{S} + \vect{v}^* \, , 
	\qquad
	p = p^\mathrm{S} + p^* \, ,     
\end{equation}
where $\vect{v}^*$ and $p^*$ are the complementary (image system) solution needed to satisfy the boundary conditions imposed at the membrane.
For an axisymmetric flow, a convenient solution form of the stationary Stokes equations has been given by Imai~\cite{imai73}, and can be expressed as~\cite{kim83}
\begin{equation}
  \vImcom_r = z \phi_{,r} + \psi_{,r} \, , 
  \qquad \vImcom_z = z \phi_{,z} - \phi + \psi_{,z} \, , 
  \qquad \pIm = 2 \eta \phi_{,z} \, , 
  \label{solutionForm}
\end{equation}
wherein $\phi$ and $\psi$ are two harmonic functions, satisfying the axisymmetric Laplace equation, to be determined from the underlying boundary conditions.
Far away from the singularity position, it can readily be checked that the solution form given above satisfies the governing equations stated by Eqs.~\eqref{StokesGleischungenCylind}.

We now denote by~$\phi_+$ and~$\psi_+$ the unknown harmonic functions in the upper-half space for~$z\ge 0$, and by $\phi_-$ and~$\psi_-$ the corresponding functions in the lower half-space for~$z \le 0$.
The general solution of the harmonic equations~$\Delta\phi_\pm = 0$ and~$\Delta\psi_\pm = 0$ in cylindrical coordinates for an axisymmetric problem can conveniently be expressed in terms of infinite integrals over the wavenumber~$q$, as~\cite{watson95}
\begin{subequations}\label{psi_pm_phi_pm}
 \begin{align}
  \phi_\pm (r,z) &= \int_0^\infty A_\pm (q) e^{-q|z|} J_0 (qr) \, \Intd q \, , \\
  \psi_\pm (r,z) &= \int_0^\infty B_\pm (q) e^{-q|z|} J_0 (qr) \, \Intd q \, , 
 \end{align}
\end{subequations}
where~$J_n$ denotes the $n$th order Bessel function of the first kind~\cite{abramowitz72}.
Moreover, the functions $A_\pm$ and $B_\pm$ are unknown wavenumber-dependent functions to be determined from the natural continuity of the velocity field across the membrane, together with the discontinuity of the hydrodynamic stresses, as derived from the shear and bending properties of the membrane.

By making use of Eqs.~\eqref{solutionForm} and \eqref{psi_pm_phi_pm}, and interchanging the derivative and integral operators, the solution for the velocity and pressure fields is given by
\begin{subequations}\label{fullSolutionIntegralForm}
	\begin{align}
		{v_r}_\pm (r,z) &= \frac{F}{8\pi\eta} \frac{r(z-h)}{s^3} - \int_0^\infty q \big( zA_\pm+B_\pm \big)e^{-q|z|} J_1(qr)\, \Intd q \, , \\
		{v_z}_\pm (r,z) &= \frac{F}{8\pi\eta} \left( \frac{2}{s}-\frac{r^2}{s^3} \right) - \int_0^\infty \big( (1+ q|z|)A_\pm \pm q B_\pm \big) e^{-q|z|} J_0(qr) \, \Intd q  \, , \\
		p_\pm     (r,z) &= \frac{F}{4\pi} \frac{z-h}{s^3} \ \mp 2\eta  \int_0^\infty  q A_\pm e^{-q|z|} J_0 (qr) \, \Intd q \, . 
	\end{align}
\end{subequations}

It worth mentioning that, since the problem is two dimensional, the equations of fluid motion can also be expressed in terms of the stream function~$\Psi_\pm$.
Accordingly, the solution of the flow problem could be reduced to the search of a single scalar function instead of simultaneously solving for the velocity and pressure fields.
The stream functions in the upper- and lower-half spaces can be presented in the following form
\begin{equation}
	\Psi_\pm (r,z) = \frac{F}{8\pi\eta} \frac{r^2}{s} 
	 -r \int_0^\infty \left( \left( |z|+\frac{1}{q} \right) A_\pm \pm B_\pm \right) e^{-q|z|} J_1 (qr) \, \Intd q \, ,
\end{equation}
where the first term corresponds to the Stokeslet solution, whereas the integral term corresponds to the image system solution.
Accordingly, the radial and axial components of the flow velocity can, respectively, be calculated as
\begin{equation}
	{v_r}_\pm = -\frac{1}{r} \frac{\partial \Psi_\pm}{\partial z} \, ,  \qquad 
	{v_z}_\pm = \frac{1}{r} \frac{\partial \Psi_\pm}{\partial r} \, .
\end{equation}

\subsection{Derivation of the mixed boundary value problem}

Having presented the general form solution of the axisymmetric problem, we next determine the expressions of the unknown functions~$A_\pm$ and $B_\pm$ from the underlying boundary conditions.
In the following, we will demonstrate that the solution of the elastohydrodynamic problem can be formulated as a mixed boundary value problem, which can then be reduced into a system of dual integral equations.

The continuity of the radial and axial velocity components at the plane of reference yields $B_+ = B_-$ and $-A_+ - q B_+ = -A_- + q B_- $.
As a result, it follows that $B_\pm = (A_- - A_+)/(2q)$.

We now define, for convenience, the wavenumber-dependent functions $f_\mathrm{S}(q) := q(A_+(q) - A_-(q))$ and $f_\mathrm{B}(q):= q(A_+(q) + A_-(q))$.
We will show in the sequel that $f_\mathrm{S}(q)$ and $f_\mathrm{B}(q)$ are in fact functions associated with the shear and bending deformation modes, respectively.

On the one hand, the tangential and normal tractions across the membrane are continuous for $r > R$.
This results into the following integral equations for the outer boundary
\begin{subequations}\label{stressContinuity}
 \begin{align}
  \int_0^\infty  f_\mathrm{S}(q) J_1 (qr) \, \Intd q &= 0 \qquad\qquad (r > R) \, , \\
  \int_0^\infty  f_\mathrm{B}(q) J_0 (qr) \, \Intd q &= 0 \qquad\qquad (r > R) \, . 
 \end{align}
\end{subequations}

On the other hand, membrane resistance toward shear and bending introduces jumps in the traction vector across the membrane for $r < R$.
The traction jump equations given by Eqs.~\eqref{tractionJumpEqs}, are expressed in cylindrical coordinates as
\begin{subequations}\label{tractionJumpEqsCylind}
 \begin{align}
  [\sigma_{rz}] &= -\frac{2 (1+C)}{3} \, \kS \left( u_{r,rr} + \frac{u_{r,r}}{r} - \frac{u_r}{r^2} \right) \, , \\
  [\sigma_{zz}] &= \kB \left( u_{z,rrrr} + \frac{2}{r} \, u_{z, rrr} - \frac{u_{z, rr}}{r^2} + \frac{u_{z,r}}{r^3} \right) \, ,
 \end{align}
\end{subequations}
wherein $u_r$ and~$u_z$ are the radial and axial displacements of the material points of the membrane relative to their initial positions in the undeformed planar state.
In an axisymmetric problem, the membrane displacement field is a function of the radial distance only.

In order to accomplish a closure of the present elastohydrodynamic problem, we require a relationship between the velocity and displacement fields.
For that purpose, we assume, for simplicity, a no-slip boundary condition at the undisplaced membrane.
Accordingly, the velocity of the fluid at $z=0$ is supposed to be identical to that of the displaced material points of the membrane.
The effect of partial slip can be explored in future studies.
Mathematically, the no-slip condition reads~\cite{bickel06, bickel07}
\begin{equation}
	\frac{\partial}{\partial t} \, \vect{u} (r,t) = \vect{v}(r,z,t) |_{z=0}  \, . 
	\label{no-slip-realSpace}
\end{equation}

If the membrane undergoes a large deformation, the no-slip boundary condition stated above should rather be applied at the displaced membrane, see for instance Refs.~\onlinecite{sekimoto93, weekley06, salez15, saintyves16, rallabandi17, daddi18stone, rallabandi18}.
Since we restrict our attention here to the system behavior in the small deformation regime such that $|\vect{u}| \ll a \ll h$, applying the no-slip condition at the undisplaced plane of the membrane should be sufficient for our investigations.

The resulting mathematical problem can conveniently be solved using Fourier transforms in time. 
Here, we employ the usual convention of a negative exponent for the forward Fourier transform.
Transforming Eq.~\eqref{no-slip-realSpace} into Fourier domain yields
\begin{equation}
	\vect{u}(r,\omega) = \left. \frac{\vect{v} (r,z,\omega)}{i\omega} \right|_{z=0} \, . 
	\label{relationBetweenVundU}
\end{equation}

By making use of Eq.~\eqref{relationBetweenVundU}, the traction jump equations across the elastic membrane given by Eqs.~\eqref{tractionJumpEqsCylind} can be expressed as
\begin{subequations}\label{tractionJumpFinal}
 \begin{align}
  [v_{r,z}] &= \left. 4i\alpha  \left( v_{r,rr} + \frac{v_{r,r}}{r} - \frac{v_r}{r^2} \right) \right|_{z=0} \, , \\
  \left[ -\frac{p}{\eta} \right] &= \left. -4i\alphaB^3 \left( v_{z,rrrr} + \frac{2 v_{z, rrr}}{r} - \frac{v_{z, rr}}{r^2} + \frac{v_{z,r}}{r^3} \right) \right|_{z=0} \, ,
 \end{align}
\end{subequations}
wherein
\begin{equation}
	\alpha := \frac{\kS}{3 B \eta \omega} \, , \qquad
	\alphaB := \left( \frac{\kB}{4\eta \omega}\right)^{1/3} \, , 
\end{equation}
are characteristic length scales for shear and bending, respectively, as previously defined in earlier works \cite{daddi16, daddi16b, daddi16c}.
Moreover, $B:=2/(1+C)$ is a dimensionless number associated with the Skalak model.

Upon substitution of the general solution for the velocity and pressure fields given by Eqs.~\eqref{fullSolutionIntegralForm} into Eq.~\eqref{tractionJumpFinal}, the following integral equations for the inner boundary are obtained
\begin{subequations}\label{integralGleischungen}
 \begin{align}
  \int_0^\infty (1-i\alpha q) f_\mathrm{S}(q) J_1 (qr) \, \Intd q &= g_\mathrm{S} (r) \qquad\qquad (0 < r < R) \, , \label{Equation_1} \\
  \int_0^\infty (1-i \alphaB^3 q^3) f_\mathrm{B}(q) J_0 (qr) \, \Intd q &= g_\mathrm{B} (r) \qquad\qquad (0 < r < R) \, . \label{Equation_2}
 \end{align}
\label{stressDiscontinuity}
\end{subequations}
Here, we have defined, for convenience, the frequency-dependent radial functions
\begin{subequations}
	\begin{align}
		g_\mathrm{S} (r) &= -\frac{F}{4\pi\eta} \frac{3i\alpha h r (4h^2-r^2)}{(h^2 + r^2)^{7/2}} \, , \label{gS} \\
		g_\mathrm{B} (r) &= - \frac{F}{4\pi\eta} \frac{9i\alphaB^3 (16h^6-72h^4 r^2 + 18h^2 r^4 +r^6)}{(h^2+r^2)^{11/2}} \, , \label{gB}
	\end{align}
\end{subequations}
where, the subscripts S and B denote shear and bending, respectively.

\subsection{Solution for an infinite membrane}

Before proceeding with the solution of the flow problem near a finite-sized membrane, we first recall the solution for the axisymmetric Green's function near an infinitely-extended elastic membrane.
This solution has been derived earlier in closed form by some of us, by means of a two-dimensional Fourier transform technique~\cite{bickel07}, see, for instance, Ref.~\onlinecite{daddi16} for more details regarding the derivation.
In the present framework, the solution for an infinite membrane can be recovered by taking the limit in Eqs.~\eqref{integralGleischungen} as $R$ goes to infinity.
Accordingly, the shear- and bending-related functions~$f_\mathrm{S}(q)$ and~$f_\mathrm{B}(q)$ can readily be obtained by performing the inverse Hankel transforms \cite{ditkin65}, to obtain
\begin{subequations} \label{f_solutionRInf}
 \begin{align}
  f_\mathrm{S}(q) &= -\frac{F}{4\pi\eta} \frac{i\alpha h q^3 }{1-i\alpha q} \, e^{-qh} \, , 
  \label{f_S_solutionRInf} \\
  f_\mathrm{B}(q) &= -\frac{F}{4\pi\eta} \frac{i \alphaB^3 q^4 (1+qh) }{1-i\alphaB^3 q^3} \, e^{-qh} \, .
  \label{f_B_solutionRInf}
 \end{align}
\end{subequations}
Using the relations $A_\pm (q) = (f_\mathrm{B}(q)\pm f_\mathrm{S}(q))/(2q)$ and $B_\pm (q) = -f_\mathrm{S}(q)/(2q^2)$, it follows that
\begin{subequations}
	\begin{align}
		A_\pm &= -\frac{i q^2 F}{8\pi\eta} 
				\left( \frac{\alpha h}{1-i\alpha q}
		        \pm \frac{\alphaB^3 q (1+qh) }{1-i\alphaB^3 q^3} \right) e^{-qh} \, , \\
		B_\pm &= \frac{F}{8\pi\eta} \frac{i\alpha h q }{1-i\alpha q} \, e^{-qh} \, ,
	\end{align}
\end{subequations}
which are in full agreement with Ref.~\onlinecite{daddi16}.
As~$\alpha$ and $\alphaB$ are both taken to infinity (corresponding to a membrane with infinite shear and bending moduli, or to a vanishing actuation frequency), these expressions reduce to the well-known solution by Blake~\cite{blake71}, for a Stokeslet acting normal to an infinitely-extended planar hard wall.

\section{Solution of the mixed boundary value problem}
\label{sec:solutionMixedBoundaryValueProblem}

Due to the decoupled nature of the shear and bending deformation modes, it appears to be more convenient to solve the elastohydrodynamic problem by considering the effects of shear and bending independently.
The overall flow field is the superposition of the individual flow fields resulting from these two deformation modes, as long as the membrane thickness is neglected~\cite{barthes02, barthes16}.
Nevertheless, this decoupling behavior only occurs near a single planar elastic membrane.
A nonlinear coupling between shear and bending has been observed for curved elastic interfaces~\cite{daddi17b, daddi17c, daddi17d, daddi18acta}, or for two closely coupled~\cite{auth07} or warped~\cite{Kosmrlj14} fluctuating membranes.

\subsection{Shear contribution}

For an idealized membrane with pure shear elasticity, the corresponding dual integral equations reads
\begin{subequations}\label{Eq_inner_UND_outer}
	\begin{align}
	 \int_0^\infty (1-i\alpha q) f_\mathrm{S}(q) J_1 (qr) \, \Intd q &= g_\mathrm{S}(r) \qquad\qquad (0<r<R) \, , \label{Eq_inner} \\ 
	 \int_0^\infty f_\mathrm{S}(q) J_1 (qr) \, \Intd q &= 0 \qquad\qquad\quad\,\,\,\, (r>R) \, , \label{Eq_outer}
	\end{align}
\end{subequations}
where $g_\mathrm{S}(r)$ is the radially-symmetric function given explicitly by Eq.~\eqref{gS}.
The goal is to solve for the unknown function~$f_\mathrm{S}(q)$ such that the dual integral equations on both the inner and outer boundaries are satisfied.

To reduce the order of the Bessel function and bring the equations into a more familiar form, we multiply both members by~$r$ and differentiate with respect to~$r$ afterwards.
Consequently, the dual integral equation can be rewritten as
\begin{subequations}\label{Eq_inner_UND_outer_1}
	\begin{align}
	 \int_0^\infty (1-i\alpha q) q f_\mathrm{S}(q) J_0 (qr) \, \Intd q &= w(r) \qquad\qquad (0<r<R) \, ,  \label{Eq_inner_1} \\ 
	 \int_0^\infty q               f_\mathrm{S}(q) J_0 (qr) \, \Intd q &= 0 \qquad\qquad\quad\,\,\, (r>R) \, , \label{Eq_outer_1}
	\end{align}
\end{subequations}
where we have defined 
\begin{equation}
 w(r):=\frac{1}{r} \frac{\Intd}{\Intd r} \big( r g_\mathrm{S}(r) \big) = -\frac{F}{4\pi\eta} \frac{3i\alpha h (8h^4-24h^2 r^2 + 3r^4)}{(h^2+r^2)^{9/2}} \, . \notag
\end{equation}

The solution of the resulting dual integral equations can be obtained following the resolution recipes given by Sneddon~\cite{sneddon66}.
The starting point consists of writing the solution in the following integral form
\begin{equation}
 	f_\mathrm{S}(q) = \frac{1}{q} \int_0^R \chi_\mathrm{S} (t) \sin (qt) \, \Intd t \, ,  \label{functionF}
\end{equation}
where~$\chi_\mathrm{S} (t)$ is an unknown function to be determined. 
This choice is motivated by the fact that the integral equation on the outer boundary is automatically satisfied.
In fact, by substituting the form solution given above into Eq.~\eqref{Eq_outer_1}, and interchanging the order of integration, one obtains
\begin{equation}
 \int_0^R \chi_\mathrm{S}(t) \, \Intd t \int_0^\infty J_0(qr) \sin (qt) \, \Intd q = 0 \qquad\qquad (t < R < r) \, ,
\end{equation}
after making use of the identity~\cite{abramowitz72}
\begin{equation}
 \int_0^\infty J_0 (qr) \sin(qt) \, \Intd q = \frac{H(t-r)}{(t^2-r^2)^{1/2}} \, ,  \label{integralWithHeaviside_sin}
\end{equation}
where $H(\cdot)$ denotes the Heaviside function.

Substituting the form solution given by Eq.~\eqref{functionF} into Eq.~\eqref{Eq_inner_1} associated with the inner boundary, and interchanging the order of integration, yields
\begin{equation}
 \int_0^R \chi_\mathrm{S}(t) \, \Intd t \int_0^\infty (1-i\alpha q) J_0(qr) \sin(qt) \, \Intd q = w(r)
 \qquad\qquad (0<r<R) \, . \label{integralEqShear}
\end{equation}

On the one hand, it follows from Eq.~\eqref{integralWithHeaviside_sin} that
\begin{equation}
 \int_0^R \chi_\mathrm{S}(t) \, \Intd t \int_0^\infty  J_0(qr) \sin(qt) \, \Intd q = \int_r^R \frac{\chi_\mathrm{S} (t) \, \Intd t}{(t^2-r^2)^{1/2}} \, . \label{firstTerm}
\end{equation}

On the other hand, an integration by parts yields
\begin{equation}
  \int_0^\infty J_0(qr) \, \Intd q \int_0^R q\chi_\mathrm{S}(t) \sin(qt) \, \Intd t 
  = \frac{\chi_\mathrm{S}(0)}{r}  + \int_0^r \frac{\chi_\mathrm{S} '(t) \, \Intd t}{(r^2-t^2)^{1/2}} \, ,
 \label{secondTerm}
\end{equation}
upon making use of the identity~\cite{abramowitz72}
\begin{equation}
 \int_0^\infty J_0 (qr) \cos(qt) \, \Intd q = \frac{H(r-t)}{(r^2-t^2)^{1/2}} \, . \label{integralWithHeaviside_cos}
\end{equation}

By inserting Eqs.~\eqref{firstTerm} and \eqref{secondTerm} into the integral equation given by Eq.~\eqref{integralEqShear}, one gets
\begin{equation}
 \int_r^R \frac{\chi_\mathrm{S} (t) \, \Intd t}{(t^2-r^2)^{1/2}} -i\alpha \left( \frac{\chi_\mathrm{S}(0)}{r} + \int_0^r \frac{\chi_\mathrm{S} '(t) \, \Intd t}{(r^2-t^2)^{1/2}} \right) = w(r)  
 \qquad\qquad
 (0<r<R)  \, . \label{integralEqShearTmp}
\end{equation}

After some rearrangements and mathematical manipulations, the latter equation can be presented in the form of an inhomogeneous Fredholm integral equation of the second kind with a logarithmic kernel.
Specifically,
\begin{equation}
 \chi_\mathrm{S}(s) = \frac{12 F h^2}{\pi^2\eta} \frac{s(h^2-s^2)}{(h^2+s^2)^4} + \frac{1}{i\pi\alpha} \int_0^R \chi_\mathrm{S}(t) \ln \left| \frac{s+t}{s-t} \right| \, \Intd t 
 \qquad\qquad (0<s<R) \, .
 \label{fredholmEqn}
\end{equation}

Further derivation details are contained in Appendix~\ref{appendix:derivationDetails}.
Remarkably, the solution at the origin is trivial by noting that~$\chi_\mathrm{S}(0)=0$.

Due to the somewhat complex nature of the resulting integral equation at hand, an analytical solution is far from being trivial.
Therefore, recourse to a numerical resolution approach is necessary.
For that purpose, we express the solution as a finite series of terms in powers of~$s$ as
\begin{equation}
 \chi_\mathrm{S}(s) = \sum_{n=1}^N c_n s^n
 \qquad\qquad (0 < s < R) \, , 
  \label{chiShear}
\end{equation}
and sample the result at Chebyshev nodes.
These correspond to the roots of the Chebyshev polynomial~\cite{mason02} of the first kind of degree $n$.
We then perform the integration analytically and solve the resulting finite linear system of equations for the unknown complex coefficients~$c_n$.
Inserting the form series solution given by Eq.~\eqref{chiShear} into Eq.~\eqref{fredholmEqn} and interchanging between the sum and the integral operators, the resulting linear system of equations reads  
\begin{equation}
	\sum_{n=1}^{N} c_n R^n \left(  S_k^n
	- \frac{R}{i\pi\alpha} \, \phi_n(S_k) \right)
	 = \frac{12F \xi^2}{\pi^2\eta R^3} \frac{S_k(\xi^2-S_k^2)}{(\xi^2+S_k^2)^4}
	 \qquad\qquad (1 \le k \le N) \, ,  
\end{equation}
where we have defined the dimensionless parameter~$\xi=h/R \in [0,\infty)$ representing the ratio of the particle-membrane distance to the radius of the membrane.
In particular, $\xi=0$ corresponds to the infinite-size limit, whereas~$\xi\to\infty$ holds for a bulk fluid, i.e., in the absence of the membrane.
The Chebyshev nodes are mapped over the interval $[0,1]$ and are given by
\begin{equation}
	S_k = \cos^2 \left( \frac{2k-1}{4N} \, \pi \right) \in [0,1] 
	\qquad\qquad (1 \le k \le N) \, .
	\label{ChebyshevNodes}
\end{equation}

Moreover, the series functions~$\phi_n(S)$ are defined by 
\begin{equation}
	\phi_n(S) = \int_0^1 T^n \ln \left| \frac{S+T}{S-T} \right| \, \Intd T 
	\qquad\qquad (0<S<1) \, , \label{phi_n_definition}
\end{equation}
the expressions of which are explicitly given by
\begin{equation}
	\phi_n (S) = 
	\begin{cases}
			\frac{1}{n+1} \left( \sum_{k=0}^{\frac{n}{2}-1} \frac{S^{n-(2k+1)} }{k+1} + (1+S^{n+1})\ln (1+S) 
			- 2S^{n+1} \ln S - (1-S^{n+1}) \ln (1-S) \right)
	 & \text{if $n$ is even} \, , \\
		\frac{1}{n+1} \left( \sum_{k=0}^{\frac{n-1}{2}} \frac{2S^{n-2k}}{2k+1} + (1-S^{n+1}) \ln \left( \frac{1+S}{1-S} \right) \right) & \text{if $n$ is odd} \, .
	\end{cases} \notag
\end{equation}	

The latter result can readily be shown by recurrence over~$n$. 
We further mention that the series~$\phi_n (S)$ can also be expressed for arbitrary parity of~$n$ in terms of the Lerch transcendent function~\cite{abkarian07} (which is implemented in computer algebra systems such as Mathematica or Maple as LerchPhi).
Specifically,
\begin{equation}
	\phi_n(S) = \ln \left( \frac{1+S}{1-S} \right) + S \left( \frac{2}{n}-i\pi S^n - \Phi \left( \frac{1}{S^2}, 1, \frac{n}{2} \right) \right)   \, . \notag 
\end{equation}

The expression of $\phi_n(S)$ for $n \le 10$ are provided for convenience in Tab.~\ref{Table_phi_n}.

\def\arraystretch{1.5}
\begin{table*}
  		\centering
    	\begin{tabular}{|c|c|}
    		\hline
    		$n$ & $\phi_n (S)$ \\
    		\hline\hline
    		0 & $-2S\ln S + (1+S)\ln(1+S)-(1-S)\ln(1-S) $ \\
    		\hline
    		1 & $S + \frac{1-S^2}{2} \ln \left( \frac{1+S}{1-S} \right) $ \\
    		\hline
    		2 & $\frac{S}{3} \left( 1-2S^2\ln S \right)
    		+\frac{1+S^3}{3} \ln (1+S) - \frac{1-S^3}{3} \ln (1-S) $ \\
    		\hline
    		3 & $\frac{S}{6}+\frac{S^3}{2} +\frac{1-S^4}{4} \ln \left( \frac{1+S}{1-S} \right) $ \\
    		\hline
    		4 & $\frac{S}{10}+\frac{S^3}{5}-\frac{2}{5} \, S^5 \ln S
    		+\frac{1+S^5}{5} \, \ln(1+S) - \frac{1-S^5}{5} \, \ln (1-S) $ \\
    		\hline
    		5 & $\frac{S}{15}+\frac{S^3}{9}+\frac{S^5}{3}
    		+\frac{1-S^6}{6} \, \ln \left( \frac{1+S}{1-S} \right) $ \\
    		\hline
    		6 & $\frac{S}{21}+\frac{S^3}{14}+\frac{S^5}{7} - \frac{2}{7} \, S^7 \ln S
    		+\frac{1+S^7}{7} \, \ln(1+S) - \frac{1-S^7}{7} \, \ln(1-S) $ \\
    		\hline
    		7 & $\frac{S}{28}+\frac{S^3}{20}+\frac{S^5}{12}+\frac{S^7}{4}
    		+\frac{1-S^8}{8} \, \ln \left( \frac{1+S}{1-S} \right) $ \\
    		\hline
    		8 & $\frac{S}{36}+\frac{S^3}{27}+\frac{S^5}{18}+\frac{S^7}{9}-\frac{2}{9} \, S^9\ln S + \frac{1+S^9}{9} \, \ln(1+S) - \frac{1-S^9}{9} \, \ln(1-S) $ \\
    		\hline
    		9 & $\frac{S}{45}+\frac{S^3}{35}+\frac{S^5}{25}+\frac{S^7}{15}+\frac{S^9}{5}
    		+ \frac{1-S^{10}}{10} \, \ln \left( \frac{1+S}{1-S} \right) $ \\
    		\hline
    		10 & $\frac{S}{55}+\frac{S^3}{44}+\frac{S^5}{33}+\frac{S^7}{22}+\frac{S^9}{11}
    		-\frac{2}{11} \, S^{11} \ln S 
    		+\frac{1+S^{11}}{11} \, \ln(1+S) - \frac{1-S^{11}}{11} \, \ln(1-S) $ \\
    		\hline
    	\end{tabular}
    \caption{Analytical expressions of the first eleven terms of the series functions~$\phi_n(S)$ defined in the main text by Eq.~\eqref{phi_n_definition}.
    Here, $S$ corresponds a Chebyshev node mapped over the interval $[0,1]$ as given by Eq.~\eqref{ChebyshevNodes}.}
  \label{Table_phi_n}
\end{table*}

\subsection{Bending contribution}

We next consider the bending-related contribution to the flow field and search for solution of the dual integral equations
\begin{subequations}
	\begin{align}
	 \int_0^\infty (1-i\alphaB^3 q^3) f_\mathrm{B}(q) J_0 (qr) \, \Intd q &= g_\mathrm{B}(r) \qquad\qquad (0<r<R) , \label{Eq_inner_Bending} \\ 
	 \int_0^\infty f_\mathrm{B}(q) J_0 (qr) \, \Intd q &= 0 \,\quad\qquad\qquad\,\,\,\, (r>R) \, , \label{Eq_outer_Bending}
	\end{align}
\end{subequations}
where~$f_\mathrm{B}(q)$ is the unknown function, and $g_\mathrm{B}(r)$ is the known radially-symmetric function given by Eq.~\eqref{gB}.
The dual integral equations have the familiar form encountered in various mixed boundary value problems where the kernel is expressed in terms of the zeroth-order Bessel function.

Similarly, we search a solution of the integral form
\begin{equation}
 f_\mathrm{B} (q) = \int_0^R \chi_\mathrm{B}(t) \sin (qt) \, \Intd t \, , \label{funtionF0}
\end{equation}
where we assume that~$\chi_\mathrm{B}(0) = \chi_\mathrm{B}(R)=0$.
This form solution satisfies the integral equation on the outer boundary after making use of Eq.~\eqref{integralWithHeaviside_sin}.
Performing three successive integrations by parts yields
\begin{equation}\label{secondTermBending}
    	\int_0^\infty J_0(qr) \, \Intd q \int_0^R q^3 \chi_\mathrm{B}(t)\sin(qt)
    	\\
    	= -  \frac{\chi_\mathrm{B}''(0)}{r} - \int_0^r \frac{\chi_\mathrm{B}'''(t) \, \Intd t}{(r^2-t^2)^{1/2}}   \, ,
\end{equation}	
where, once again, we have made use of the identity given by Eq.~\eqref{integralWithHeaviside_cos}.
We have further assumed that $\chi_\mathrm{B}'(t)$ vanishes at the domain boundary, so as to ensure convergence of the overall improper integral.

By making use of Eqs.~\eqref{integralWithHeaviside_sin} and \eqref{secondTermBending}, the integral equation on the inner boundary becomes
\begin{equation}
		 \int_r^R \frac{\chi_\mathrm{B}(t) \, \Intd t}{(t^2-r^2)^{1/2}} \\
		 + i \alphaB^3 \left(  \frac{\chi_\mathrm{B}''(0)}{r} +  \int_0^r \frac{\chi_\mathrm{B}'''(t) \, \Intd t}{(r^2-t^2)^{1/2}} \right) = g_\mathrm{B}(r)
		 \qquad\qquad (0<r<R)  \, .
\end{equation}

The latter can be written in the form of an inhomogeneous Fredholm equation of the second kind by using the same resolution procedure described in Appendix~\ref{appendix:derivationDetails}.
Integrating the resulting equation with respect to~$s$ twice yields
\begin{equation}
	\chi_\mathrm{B}(s) = \frac{4F h^3}{\pi^2 \eta} \frac{s}{\left(s^2+h^2\right)^3} + \frac{b_1 s}{R} + b_2
	- \frac{1}{i\pi\alphaB^3} \int_0^R \chi_\mathrm{B} (t) \left( K(t,s) + \frac{b_3 s}{R} + b_4 \right) \, \Intd t 
	\qquad\qquad (0<s<R) \, , 
	 \label{fredholmEqn_Bending}
\end{equation}
where $b_1, b_2, b_3$, and $b_4$ are four unknown integration constants to be determined from the imposed boundary conditions, namely $\chi_\mathrm{B}(0)=\chi_\mathrm{B}(R)=0$.
The function appearing in the kernel of the integral is given by
\begin{equation}
	K(t,s) = \frac{(t+s)^2}{2} \, \ln(s+t) - \frac{(t-s)^2}{2} \, \ln (|s-t|) -ts\left( 1+2\ln(2t) \right) \, .
\end{equation}

Since an analytical solution is far from being intuitive, we attempt a numerical evaluation of the integral equation by considering a solution of the form
\begin{equation}
	\chi_\mathrm{B}(s) = \sum_{n=1}^{N} d_n s^n 
	\qquad\qquad (0<s<R) \, , 
	 \label{lambdaBending}
\end{equation}
and sample the result at Chebyshev nodes, in the same way as previously done for the shear-related part.
After some rearrangement, we obtain
\begin{equation}
	\sum_{n=1}^{N} d_n R^n \left( S_k^n + \frac{R^3}{i\pi\alphaB^3} \left( \rho_n(S_k) + \frac{b_3 S_k+b_4}{n+1} \right)	\right) = \frac{4F \xi^3}{\pi^2\eta R^2} \frac{S_k}{(S_k^2+\xi^2)^3} + b_1 S_k + b_2 
	\qquad\qquad (1 \le k \le N) \, ,
\end{equation}
where we have defined the series functions
\begin{equation}
	\rho_n(S) = \int_0^1 T^n K(T,S) \, \Intd T 
	\qquad\qquad (0<S<1) \, , 
\end{equation}
the general term of which is given by
\begin{equation}
\rho_n (S) = R_n(S) -S \left( \frac{1}{n+3} + \frac{2\ln 2}{n+2} \right) +
\frac{1}{(n+1)(n+2)(n+3)} \left(Q_n+
\begin{cases}
	-2S^{n+3} \ln S &  \text{if $n$ is even}  \\
	0 &  \text{if $n$ is odd}  \\
\end{cases} \right) \, , \label{rho_n}
\end{equation}
where	
\begin{subequations}
	\begin{align}
			R_n (S) &= \frac{1}{(n+1)(n+2)(n+3)} \sum_{k=0}^{n} \tfrac{(k+1)(k+2)}{2}  \left( (-1)^{k+n} f_+(S) - f_-(S) \right) S^{n-k}   \, , \\
			Q_n (S) &= \sum_{k=0}^{\ceil[\big] {\frac{n}{2}-1} } \frac{S^{2k+3}}{\frac{n}{2}-k} \, , 
	\end{align}
\end{subequations}
wherein $f_\pm (S) = (1\pm S)^3 \ln (1\pm S)$.
Moreover, $\ceil[\big]{x} = \operatorname{ceil} (x)$  denotes the ceiling function, which maps a variable~$x$ to the least integer greater than or equal to $x$.
Depending on the parity of~$n$, the general term of~$R_n(S)$ and $Q_n(S)$ are, respectively, expressed by
\begin{subequations}
	\begin{align}
	R_n (S) &=
		\begin{cases}
			\frac{S}{n+2} \left( 1+\frac{S^{n+2}}{(n+1)(n+3)} \right) \ln \left(1-S^2\right)
					+\frac{1}{2} \left( \frac{S^2}{n+1} + \frac{1}{n+3} \right) \ln \left( \frac{1+S}{1-S} \right) & \text{if $n$ is even} \, , \\
			\frac{S}{n+2} \ln (1-S^2)
			 + \left( \frac{S^2}{2(n+1)}+\frac{1}{2(n+3)}-\frac{S^{n+3}}{(n+1)(n+2)(n+3)} \right) \ln \left( \frac{1+S}{1-S} \right) & \text{if $n$ is odd}  \, , 
		\end{cases} 
	\end{align}
\end{subequations}
and 
\begin{subequations}
	\begin{align}
		Q_n (S) &=
				\begin{cases}
					S^{n+1} + S^{n-1} \Phi \left( \frac{1}{S^2},1,2 \right) - S \Phi \left( \frac{1}{S^2}, 1, \frac{n}{2} +1 \right) & \text{if $n$ is even} \, , \\
					S^{n+3} \ln \left( \frac{1+S}{1-S} \right)
					-S^3 \Phi \left( S^2,1,-\frac{n}{2} \right) & \text{if $n$ is odd} \, .
				\end{cases} 
	\end{align}
\end{subequations}

Expressions for $\rho_n(S)$ become rather lengthy and cumbersome as~$n$ gets larger, and thus have not been provided here.
These can more advantageously be obtained using computer algebra systems.

Finally, it is worth mentioning that the solution given by Eqs.~\eqref{f_solutionRInf} for~$R\to\infty$ can also be recovered from the resulting integral equations of the mixed boundary problem.
The calculation details have been relegated to Appendix~\ref{appendix:recoveryOFSolutionAsRgoesToInfty}.

\section{Hydrodynamic mobility}
\label{sec:hydrodynamicMobility}

The calculation of the flow field induced by a point-force acting near the elastic membrane can be employed to assess the effect of the interface on the slow motion of a suspended particle moving in its vicinity.
This membrane-induced effect is quantified by the hydrodynamic mobility functions, which are tensorial quantities that bridge between the velocity of a sedimenting particle and the force exerted on its surface~\cite{batchelor76}. 
In a bulk fluid, the particle mobility is isotropic and may be obtained from by the Stokes resistance formula as $\mu_{\alpha\beta} = \mu_0 \delta_{\alpha\beta}$, where $\mu_0 = 1/(6\pi\eta a)$ and $\delta_{\alpha\beta}$ denotes the Kronecker symbol. 
The presence of the elastic membrane introduces a correction to the hydrodynamic mobility that is dependent not only on the particle size and the viscosity of the suspending fluid, but also on the distance separating the particle from the membrane, as well as on the forcing frequency in the system.

In the following, we present analytical expressions for the frequency-dependent mobility of a spherical particle translating perpendicular to a finite-sized membrane possessing either shear elasticity or bending rigidities.
The mobility correction near a membrane endowed simultaneously with both shear and bending deformation modes can be obtained by linear superposition of the shear- and bending-induced corrections as obtained independently.
Again, this is true only for a single planar membrane where a decoupling between shear and bending deformation modes exists~\cite{daddi16b}.

\subsection{Shear contribution}

For an idealized elastic membrane with only-shear resistance, such as that of a red blood cell, the bending-related part in the solution of the flow problem vanishes.
Thus, $A_\pm = \pm f_\mathrm{S}(q)/(2q)$ and $B_\pm = -f_\mathrm{S}(q)/(2q^2)$.
In the point-particle approximation, the scaled correction to the particle mobility at leading order for the motion perpendicular to the finite-sized membrane can directly be calculated by evaluating the correction to the axial component of the flow field at the particle position.
Specifically,
\begin{equation}
  \frac{\Delta \mu_\mathrm{S}}{\mu_0} := 
  \left. \frac{F^{-1}}{\mu_0} \lim_{(r,z)\to(0,h)} v_z^{*} \right|_{\alphaB=0}
  = -3\pi\eta a h F^{-1} \int_0^\infty f_\mathrm{S}(q) \, e^{-qh} \, \Intd q \, . \label{mobilityCorrection_def}
\end{equation}

As shown in the previous section, an analytical solution for~$\chi_\mathrm{S}(t)$ is not available due to the complicated nature of the resulting Fredholm integral equation.
Therefore, a numerical technique has been employed in order to overcome this difficulty.
Substituting the form series solution given by Eq.~\eqref{chiShear} into Eq.~\eqref{functionF} and interchanging between the sum and the integral operator, yields
\begin{equation}
 	f_\mathrm{S}(q) = \frac{1}{q} \sum_{n=1}^N c_n R^{n+1} \int_0^1  T^n \sin (qR T) \, \Intd T \, . \label{f_S_series}
\end{equation}

By inserting the latter equation into Eq.~\eqref{mobilityCorrection_def} and performing integration with respect to the variable~$q$, the shear-related part in the correction to the frequency-dependent mobility can be obtained in a scaled form as
\begin{equation}
 \begin{split}
  \frac{\Delta \mu_\mathrm{S}}{\mu_0} = -3\pi\eta a h F^{-1} \sum_{n=1}^N c_n R^{n+1} \psi_n (\xi) \, , \label{deltaMu_Shear}
 \end{split}
\end{equation}
where we have defined the series functions
\begin{equation}
	\psi_n (\xi) = \int_0^1 T^n \arctan \left( \frac{T}{\xi} \right) \, \Intd T 
	=  \frac{1}{n+1} 
		\left( \frac{\pi}{2} \left( 1 + \frac{\xi^{n+1}}{\sin \left( \frac{n\pi}{2} \right)} \right)-\arctan \xi + \frac{\xi}{2} \, \Phi \left( -\xi^2,1,-\frac{n}{2} \right) \right) 
	\, , \label{psiDef}
\end{equation}
which, depending on the parity of~$n$, is explicitly given by 
\begin{equation}
	\psi_n (\xi) = 
	\begin{cases}
		\frac{1}{n+1} \left( \frac{1}{2} \sum_{k=0}^{\frac{n}{2}-1} \frac{(-1)^{k+\frac{n}{2}}}{k+1} \, \xi^{n-(2k+1)} + \frac{1}{2} (-1)^{\frac{n}{2}+1} \, \xi^{n+1} \ln \left(1+\frac{1}{\xi^2}\right) + \arctan \left(\frac{1}{\xi}\right)  \right) & \text{if $n$ is even} \, ,  \\
		\frac{1}{n+1} 
		\left( \sum_{k=0}^{\frac{n-1}{2}} \frac{(-1)^{k+\frac{n+1}{2}}}{2k+1} \, \xi^{n-2k} 
		+ (1+\xi^{n+1}) \arctan \left( \frac{1}{\xi} \right) \right) & \text{if $n$ is odd}  \, .
	\end{cases}
\end{equation}

Expressions of the series functions~$\psi_n(\xi)$ have been provided for convenience in Tab.~\ref{Table_psi_n_zeta_n} for $n\le 10$.

\subsubsection*{Exact solution for infinite shear modulus}

For an infinite membrane shear modulus, or equivalently for a vanishing actuation frequency, the shear length scale~$\alpha\to\infty$.
In this situation, the solution of the integral equation associated with shear simplifies to (c.f.~Eq.~\eqref{fredholmEqn})
\begin{equation}
  \lim_{\alpha\to\infty} \chi_\mathrm{S}(t) = \frac{12F h^2}{\pi^2\eta} \frac{t(h^2-t^2)}{(h^2+t^2)^4} \, .
\end{equation}

Substituting the latter equation into Eq.~\eqref{functionF} yields
\begin{equation}
  \lim_{\alpha\to\infty} f_\mathrm{S} (q) = \frac{12 F h^2}{\pi^2\eta q}  \int_0^R \frac{t(h^2-t^2)}{(h^2+t^2)^4} \, \sin(qt) \, \Intd t \, . \label{f_S}
\end{equation}

By inserting this solution into Eq.~\eqref{mobilityCorrection_def} and performing a double integration, the scaled correction to the hydrodynamic mobility to leading order near a membrane with infinite shear reads
\begin{equation}
  \lim_{\alpha\to\infty} \frac{\Delta \mu_\mathrm{S}}{\mu_0} = -\frac{1}{8\pi(1+\xi^2)^3} \left( 3\left( 1+3 \xi^2+27\xi^4-7\xi^6 \right) \arctan \left( \frac{1}{\xi} \right) + \xi \left( 3+8\xi^2+21\xi^4 \right) \right) \frac{a}{h} \, .
  \label{deltaMu_Shear_Steady}
\end{equation}

For a large size of the membrane compared to the particle-membrane distance, such that  $\xi \ll 1$, the leading-order correction to the scaled mobility can be expanded as power series of~$\xi$ as
\begin{equation}
   \lim_{\alpha\to\infty} \frac{\Delta \mu_\mathrm{S}}{\mu_0} = \left( -\frac{3}{16} - \frac{9}{2} \, \xi^4+\frac{36}{5\pi} \, \xi^5 + 15 \, \xi^6 \right) \frac{a}{h} + \bigO (\xi^7) \, .
   \label{deltaMu_Shear_Steady_Expansion}
\end{equation}

Particularly, as $\xi\to 0$, we recover in the vanishing-frequency limit the shear-related contribution to the mobility correction for an infinitely-extended membrane~\cite{daddi16}.

\subsection{Bending contribution}

We next consider an idealized membrane with only bending resistance, such as that of a lipid membrane.
Accordingly, the shear-related part in the solution of the flow problem vanishes.
This leads to~$A_\pm = f_\mathrm{B}(q)/(2q)$ and $B_\pm = 0$.
The scaled particle mobility correction due to bending is calculated to leading order as
\begin{equation}
  \frac{\Delta \mu_\mathrm{B}}{\mu_0} 
  =  \left. \frac{F^{-1}}{\mu_0} \lim_{(r,z)\to(0,h)} v_z^{*} \right|_{\alpha=0}
  = -3\pi\eta a F^{-1} \int_0^\infty \frac{1 + qh}{q} \, f_\mathrm{B}(q) \, e^{-qh} \, \Intd q \, . \label{mobilityCorrection_def_bending}
\end{equation}

By substituting the series representation of $\chi_\mathrm{B}(t)$ given by Eq.~\eqref{lambdaBending} into Eq.~\eqref{funtionF0} and interchanging between the sum and the integral, we obtain
\begin{equation}
	f_\mathrm{B}(q) = \sum_{n=1}^{N} d_n R^{n+1} \int_0^1 T^n \sin (qRT) \, \Intd T \, . \label{f_B_series}
\end{equation}

Inserting the latter equation into \eqref{mobilityCorrection_def_bending}, and integrating with respect to~$q$, the leading-order correction to the scaled hydrodynamic mobility due to bending reads
\begin{equation}
	\frac{\Delta \mu_\mathrm{B}}{\mu_0} = -3\pi\eta a h F^{-1} 
	\sum_{n=1}^{N} d_n R^{n+1} \left( \psi_n (\xi) + \xi \,  \zeta_n (\xi) \right) \, , 
	\label{deltaMu_Bending}
\end{equation}
where we have defined 
\begin{equation}
	\zeta_n (\xi) = \int_0^1 \frac{T^{n+1}}{\xi^2 + T^2} \, \Intd T
	= (-1)^{2n+1} \left( \frac{1}{2} \, \Phi \left( -\frac{1}{\xi^2},1,\frac{n}{2} \right) - \frac{1}{n} \right) \, , \label{zetaDef}
\end{equation}
which, depending on the parity of~$n$, is given by
\begin{equation}
	\zeta_n (\xi) = \sum_{k=0}^{\ceil[\big] {\frac{n}{2}-1}} (-1)^k \frac{\xi^{2k}}{n-2k}
	+
	\begin{cases}
		(-1)^{\frac{n}{2}} \left(\frac{1}{2} \ln \left(1+\xi^2\right) - \ln\xi\right) \xi^{n} & \text{if $n$ is even} \, ,  \\
		(-1)^{\frac{n+1}{2}} \xi^n \arctan \left( \frac{1}{\xi} \right) & \text{if $n$ is odd} \, .
	\end{cases}
\end{equation}

Expressions of~$\zeta_n(\xi)$ for $n\le 10$ have been provided in Tab.~\ref{Table_psi_n_zeta_n}.

\def\arraystretch{1.5}
\begin{table*}
	\centering
	\begin{tabular}{|c|c|c|}
	\hline
	$n$ & $\psi_n (\xi)$ & $\zeta_n (\xi)$\\
	\hline\hline
	0 & $\arctan \left(\frac{1}{\xi}\right) - \frac{\xi}{2} \, \ln \left( 1+\frac{1}{\xi^2} \right)$ & $ f $ \\
	\hline
	1 & $-\frac{\xi}{2} + \frac{1+\xi^2}{2} \, \arctan \left(\frac{1}{\xi}\right) $ & $ 1-\xi \arctan \left( \frac{1}{\xi} \right) $\\
	\hline 
	2 & $-\frac{\xi}{6} + \frac{\xi^3}{6} \, \ln \left( 1+\frac{1}{\xi^2} \right) + \frac{1}{3} \, \arctan \left(\frac{1}{\xi}\right) $ & $ \frac{1}{2} - f \xi^2 $ \\
	\hline
	3 & $-\frac{\xi}{12}+\frac{\xi^3}{4} + \frac{1-\xi^4}{4} \, \arctan \left(\frac{1}{\xi}\right) $ & $ \frac{1}{3}-\xi^2 + \xi^3 \arctan \left(\frac{1}{\xi}\right) $ \\
	\hline
	4 & $-\frac{\xi}{20}+\frac{\xi^3}{10} - \frac{\xi^5}{10} \, \ln \left( 1+\frac{1}{\xi^2} \right) + \frac{1}{5} \, \arctan \left(\frac{1}{\xi}\right) $ & $ \frac{1}{4}-\frac{\xi^2}{2} + \xi^4 f $\\
	\hline
	5 & $-\frac{\xi}{30}+\frac{\xi^3}{18}-\frac{\xi^5}{6} + \frac{1+\xi^6}{6} \, \arctan \left(\frac{1}{\xi}\right) $ & $ \frac{1}{5}-\frac{\xi^2}{3}+\xi^4-\xi^5 \arctan \left( \frac{1}{\xi} \right) $\\
	\hline
	6 & $-\frac{\xi}{42}+\frac{\xi^3}{28}-\frac{\xi^5}{14}+ \frac{\xi^7}{14} \, \ln \left( 1+\frac{1}{\xi^2} \right)+\frac{1}{7} \, \arctan \left(\frac{1}{\xi}\right)  $ & $ \frac{1}{6}-\frac{\xi^2}{4}+\frac{\xi^4}{2}- f \xi^6 $ \\
	\hline
	7 & $-\frac{\xi}{56}+\frac{\xi^3}{40}-\frac{\xi^5}{24}+\frac{\xi^7}{8} + \frac{1-\xi^8}{8} \, \arctan \left(\frac{1}{\xi}\right) $ & $ \frac{1}{7}-\frac{\xi^2}{5}+\frac{\xi^4}{3}-\xi^6 + \xi^7 \arctan \left( \frac{1}{\xi} \right) $\\
	\hline
	8 & $-\frac{\xi}{72}+\frac{\xi^3}{54}-\frac{\xi^5}{36}+\frac{\xi^7}{18} - \frac{\xi^9}{18} \, \ln \left( 1+\frac{1}{\xi^2} \right) + \frac{1}{9} \, \arctan \left(\frac{1}{\xi}\right)  $ & $ \frac{1}{8}-\frac{\xi^2}{6}+\frac{\xi^4}{4} - \frac{\xi^6}{2} + f \xi^8 $\\
	\hline
	9 & $-\frac{\xi}{90}+\frac{\xi^3}{70}-\frac{\xi^5}{50}+\frac{\xi^7}{30}-\frac{\xi^9}{10} + \frac{1+\xi^{10}}{10} \, \arctan \left(\frac{1}{\xi}\right) $ & $ \frac{1}{9} - \frac{\xi^2}{7} + \frac{\xi^4}{5}-\frac{\xi^6}{3} + \xi^8 - \xi^9 \arctan \left( \frac{1}{\xi} \right) $ \\
	\hline
	10 & $-\frac{\xi}{110}+\frac{\xi^3}{88}-\frac{\xi^5}{66} + \frac{\xi^7}{44} - \frac{\xi^9}{22} + \frac{\xi^{11}}{22} \, \ln \left( 1+\frac{1}{\xi^2} \right) + \frac{1}{11} \, \arctan \left(\frac{1}{\xi}\right)  $ & $ \frac{1}{10} - \frac{\xi^2}{8} + \frac{\xi^4}{6} - \frac{\xi^6}{4} + \frac{\xi^8}{2} - f \xi^{10} $\\
	\hline	
	\end{tabular}
	\caption{Analytical expressions of the first eleven terms of the series functions $\psi_n(\xi)$ and $\zeta_n(\xi)$ defined by Eqs.~\eqref{psiDef} and \eqref{zetaDef}, respectively.
	We recall that $\xi = h/R$ and $f = \tfrac{1}{2} \, \ln(1+\xi^2)-\ln\xi$.
	} 
	\label{Table_psi_n_zeta_n}
\end{table*}

\subsubsection*{Exact solution for infinite bending modulus}

For an infinite membrane bending modulus, or equivalently in the limit of vanishing frequency, $\alphaB \to \infty$.
The solution of the flow problem in this limiting case can readily be obtained from Eq.~\eqref{fredholmEqn_Bending} as 
\begin{equation}
	\lim_{\alphaB\to\infty} \chi_\mathrm{B}(t) = \frac{4F h^3 t}{\pi^2 \eta} \left( \frac{1}{\left(t^2+h^2\right)^3} - \frac{1}{\left(R^2+h^2\right)^3}\right)
 \, , \label{limAlphaB_chiB}
\end{equation}
where the integration constants~$b_1$ and~$b_2$ have been determined by imposing the conditions required for convergence, namely~$\chi_\mathrm{B}(0)=\chi_\mathrm{B}(R)=0$.
Inserting Eq.~\eqref{limAlphaB_chiB} into Eq.~\eqref{funtionF0} yields
\begin{equation}
 	\lim_{\alphaB\to\infty} f_\mathrm{B} (q) = \frac{4F h^3 }{\pi^2 \eta} \int_0^R \left( \frac{1}{\left(t^2+h^2\right)^3} - \frac{1}{\left(R^2+h^2\right)^3}\right) t \sin (qt) \, \Intd t \, . \label{f_B}
\end{equation}

The scaled correction to the particle mobility to leading order near a membrane with infinite bending can be obtained by substituting Eq.~\eqref{f_B} into Eq.~\eqref{mobilityCorrection_def_bending}, and performing a double integration, to obtain
\begin{equation}
	\lim_{\alphaB\to\infty} \frac{\Delta \mu_\mathrm{B}}{\mu_0} = -\frac{1}{8\pi \left(1+\xi^2\right)^3}
	\left( 15\xi+40\xi^3-39\xi^5 + \left(15+45\xi^2-27\xi^4+39\xi^6\right) \arctan \left(\frac{1}{\xi}\right) \right) \frac{a}{h} \, ,
	\label{deltaMu_Bending_Steady}
\end{equation}
which, for $\xi \ll 1$, can be expanded perturbatively in~$\xi$ as
\begin{equation}
	\lim_{\alphaB\to\infty} \frac{\Delta \mu_\mathrm{B}}{\mu_0}  = \left( -\frac{15}{16} + \frac{9}{2} \, \xi^4 - 15 \, \xi^6 \right) \frac{a}{h} + \bigO \left(\xi^7\right) \, .
	\label{deltaMu_Bending_Steady_Expansion}
\end{equation}

In particular, we recover as $\xi\to 0$ the bending-related contribution to the leading-order mobility correction in the vanishing-frequency limit~\cite{daddi16}.
It is worth mentioning that this limit is identical to that obtained near a flat liquid-liquid interface of infinite surface tension, that separates two immiscible liquids having the same viscosity~\cite{lee79}.

\subsection{Comparison with the hard disk limit}

The total mobility correction near a membrane endowed simultaneously with both shear and bending is obtained by summing up the contributions stemming from these deformation modes as obtained independently.
In particular, for infinite membrane shear and bending moduli, it follows from Eqs.~\eqref{deltaMu_Shear_Steady} and~\eqref{deltaMu_Bending_Steady} that
\begin{equation}
	\lim_{\alpha, \alphaB \to \infty} \frac{\Delta\mu}{\mu_0} = -\frac{3}{4\pi} 
	\left( \frac{3\xi+8\xi^3-3\xi^5}{\left(1+\xi^2\right)^3} +3\arctan \left(\frac{1}{\xi}\right) \right) \frac{a}{h} \, , 
	\label{deltaMu_Both_Steady}
\end{equation}
which, for $\xi \ll 1$, can be expanded in power series of~$\xi$ as
\begin{equation}
	\lim_{\alpha, \alphaB \to \infty} \frac{\Delta\mu}{\mu_0} =
	\left( -\frac{9}{8} + \frac{36}{5\pi} \, \xi^5 \right)\frac{a}{h}
	+ \bigO \left(\xi^7\right) \, .
\end{equation}

We thus recognize as $\xi\to 0$ the correction factor associated with axisymmetric motion normal to an infinite planar hard wall, as first obtained by Lorentz~\cite{lorentz07}, about one century ago. 
Interestingly, the next leading-order terms with~$\xi^4$ in Eqs.~\eqref{deltaMu_Shear_Steady_Expansion} and \eqref{deltaMu_Bending_Steady_Expansion} drop  out where the resulting correction to the mobility in the vanishing frequency limit scales rather as~$\xi^5$.

Near a hard disk, the leading-order correction to the hydrodynamic mobility has been obtained by Kim~\cite{kim83}.
The latter formulated the mixed boundary value problem and obtained a formal expression for the stream function associated with the axisymmetric flow field due to a Stokeslet near a no-slip disk.
Using our notation, the leading-order correction has been obtained as
\begin{equation}
	\frac{\Delta\mu^\mathrm{Disk}}{\mu_0} = -\frac{3}{4\pi} 
	\left( \frac{\xi(3+5\xi^2)}{\left(1+\xi^2\right)^2} +3\arctan \left( \frac{1}{\xi} \right) \right) \frac{a}{h} \,\,  \stackrel{\xi \ll 1}{=} \,\,
	\left( -\frac{9}{8} + \frac{6}{5\pi} \, \xi^5 \right)\frac{a}{h}
		+ \bigO \left(\xi^7\right) \, .
	\label{deltaMu_hardDisk}
\end{equation}

Subtracting Eq.~\eqref{deltaMu_hardDisk} from Eq.~\eqref{deltaMu_Both_Steady} yields
\begin{equation}
	\lim_{\alpha, \alphaB \to \infty} \Delta\mu - \Delta\mu^\mathrm{Disk}
	= \frac{1}{\pi^2 \eta h} \frac{\xi^5}{\left(1+\xi^2\right)^3}  \, .
\end{equation}

In the vanishing-frequency limit, the particle mobility near a finite-sized membrane with both shear and bending resistances is found to be larger than that near a no-slip disk of equal size.
The difference between the two mobilities decays rapidly as fifth power of membrane size until it eventually vanishes for an infinite membrane radius.
The essential distinction between a membrane with infinite shear and bending moduli and the hard disk limit is that, while a hard disk remains stationary (the no-slip boundary condition imposed at a hard disk implies that~$\vect{v}(r, z=0)|_{r<R} = \vect{0}$), a finite-sized elastic membrane with infinite shear and bending moduli can still undergo translational motion.
This follows from the nature of the prescribed boundary conditions.
In our model membrane, shear and bending deformation modes induce, respectively, discontinuities along the tangential and normal tractions.
Thus, even in the case of infinite shear and bending moduli, the fluid velocity at the membrane does not necessarily vanish for a finite-sized membrane.
Consequently, motion near a hard disk is more restricted and thus the particle hydrodynamic mobility is significantly reduced.
For an infinitely-extended membrane, i.e., when~$\xi \to 0$, the mobility near a membrane with infinite shear and bending is found to be identical to that predicted near a no-slip wall.
The membrane in this situation remains stationary and thus explaining the fact that the hard wall limit is recovered in this limiting case.
An analogous behavior has been observed for particle motion inside a spherical elastic cavity~\cite{daddi18creeping}, where the particle mobility inside an elastic cavity with infinite membrane shear and bending has been found to be larger than that predicted inside a rigid cavity.
This deviation has been attributed to the motion of the elastic cavity whose hydrodynamic mobility vanishes in the limit of infinite radius.

The velocity of the material points constituting the membrane can be calculated by evaluating the flow velocity field at~$z=0$ as
\begin{equation}
	v_z^\mathrm{Mem} (r) := \left. v_z (r,z) \right|_{z=0}  = \frac{F}{8\pi\eta} \left( \frac{2}{\left(r^2+h^2\right)^{1/2}} - 
	\frac{r^2}{\left(r^2+h^2\right)^{3/2}}	\right)
	-\frac{1}{2} \int_0^\infty \frac{f_\mathrm{B} (q)}{q} \, J_0 (qr) \, \Intd q \, .
\end{equation}

Interestingly, the latter is only a function of the membrane bending properties. 
By considering the limiting case of infinite bending modulus, and substituting the corresponding expression of $f_\mathrm{B} (q)$ from Eq.~\eqref{f_B} in the above integral, the average membrane velocity reads
\begin{equation}
	\langle v_z^\mathrm{Mem} \rangle
	 := \frac{1}{2R} \int_0^R v_z^\mathrm{Mem} (r) \, r \, \Intd r 
	 = \frac{F}{32\pi\eta h}  \frac{\xi^4 (5+2\xi^2)}{\left(1+\xi^2\right)^3} 
	 \stackrel{\xi \to 0}{\to} \,\, 0 \,  .
\end{equation}
Here, we have made use of the identity
\begin{equation}
	\int_0^\infty \frac{\sin (qt)}{q} \, J_0(qr) \, \Intd q 
	= \frac{\pi}{2} + \left( \arctan \left( \frac{t}{\left(r^2-t^2\right)^{1/2}} \right) - \frac{\pi}{2} \right) H(r-t) \,  ,
\end{equation}

\begin{figure}
\begin{center}
\includegraphics[scale=1.]{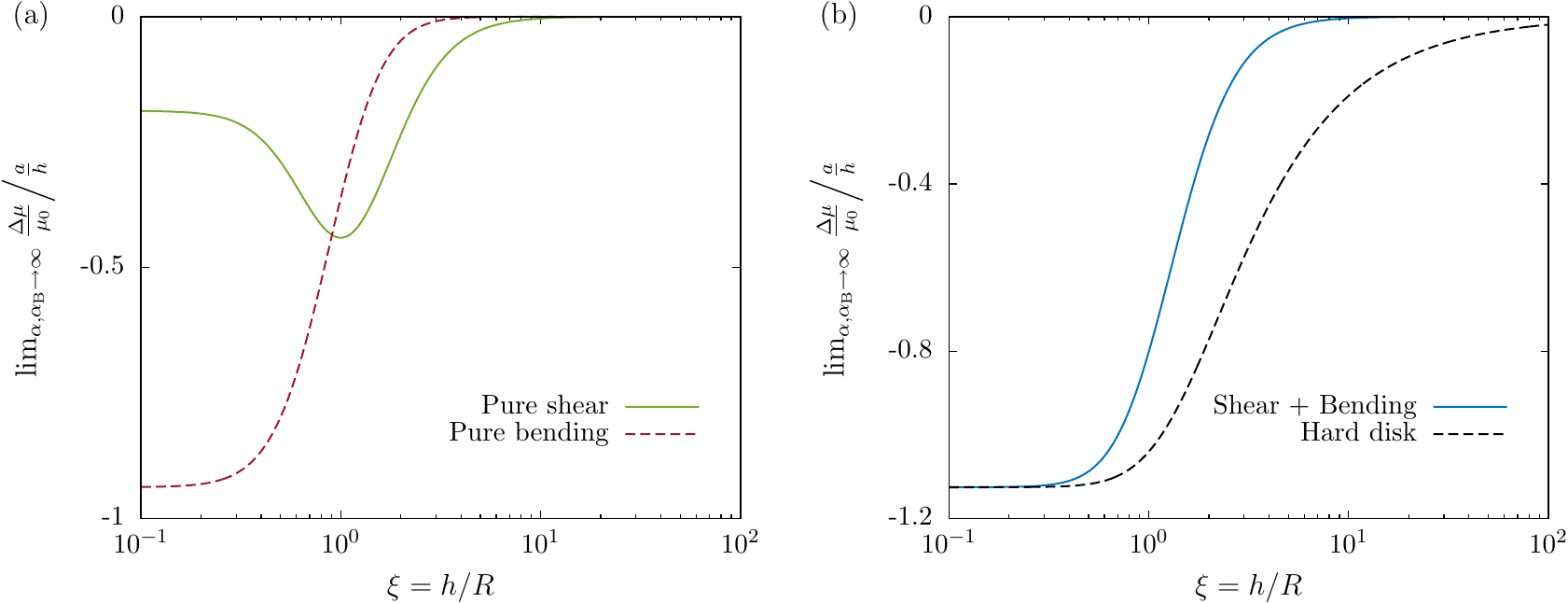}
\end{center}
\caption{(Color online) Correction factor in the hydrodynamic mobility function versus the ratio of the particle-membrane distance to the radius of the disk, in the limit of vanishing frequency.
(a) Results are shown for idealized membranes with pure shear or bending resistances, as predicted theoretically by Eqs.~\eqref{deltaMu_Shear_Steady} and \eqref{deltaMu_Bending_Steady}.
(b) Variations of the correction factor for a membrane possessing both shear and bending resistances and a hard disk with no-slip boundary conditions, as given by Eqs.~\eqref{deltaMu_Both_Steady} and \eqref{deltaMu_hardDisk}.
}
\label{MobiCorr_VanishingFreq}
\end{figure}

In Fig.~\ref{MobiCorr_VanishingFreq}~(a), we show the variations of the scaled correction to the particle mobility in the vanishing-frequency limit versus~$\xi$ for idealized membranes endowed with pure shear (green solid line) or pure bending (red dashed line).
While bending-related contribution decreases monotonically in magnitude upon increasing~$\xi$, the shear-related part shows a peak value at $\xi=1$.
The latter is attributed to the fact that the shear stresses attain their maximum value (in magnitude) when~$R=h$ and thus causing an increased resistance to the motion of the nearby particle.
Both the shear and bending-related corrections are found to be of about the same magnitude for $\xi = 1$.
Hence, the system behavior is bending dominated below this value, and shear dominated above. 
This effect is directly linked to the magnitude of stresses at the membrane, where the bending-mediated normal stresses dominate over the shear stresses when~$\xi < 1 $, and vice versa (c.f.\ Appendix~\ref{appendix:NEU} for further details).

Fig.~\ref{MobiCorr_VanishingFreq}~(b) presents a comparison between the scaled mobility correction in the vanishing-frequency limit near a membrane endowed with both shear and bending resistances (blue solid line) and the scaled mobility correction near a no-slip disk (black dashed line) as functions of the dimensionless parameter~$\xi$.
Both curves increase monotonically with~$\xi$ and eventually vanish as $\xi$ is taken to infinity where the bulk behavior is recovered.
As already mentioned, the two corrections amount to the same value only in the limit when $\xi\to 0$ corresponding to an infinitely-extended interface.
Apart from that, the correction near a hard disk is always found to be larger in magnitude compared to that near a finite-sized elastic membrane.

\begin{figure}
\begin{center}
\includegraphics[scale=1.]{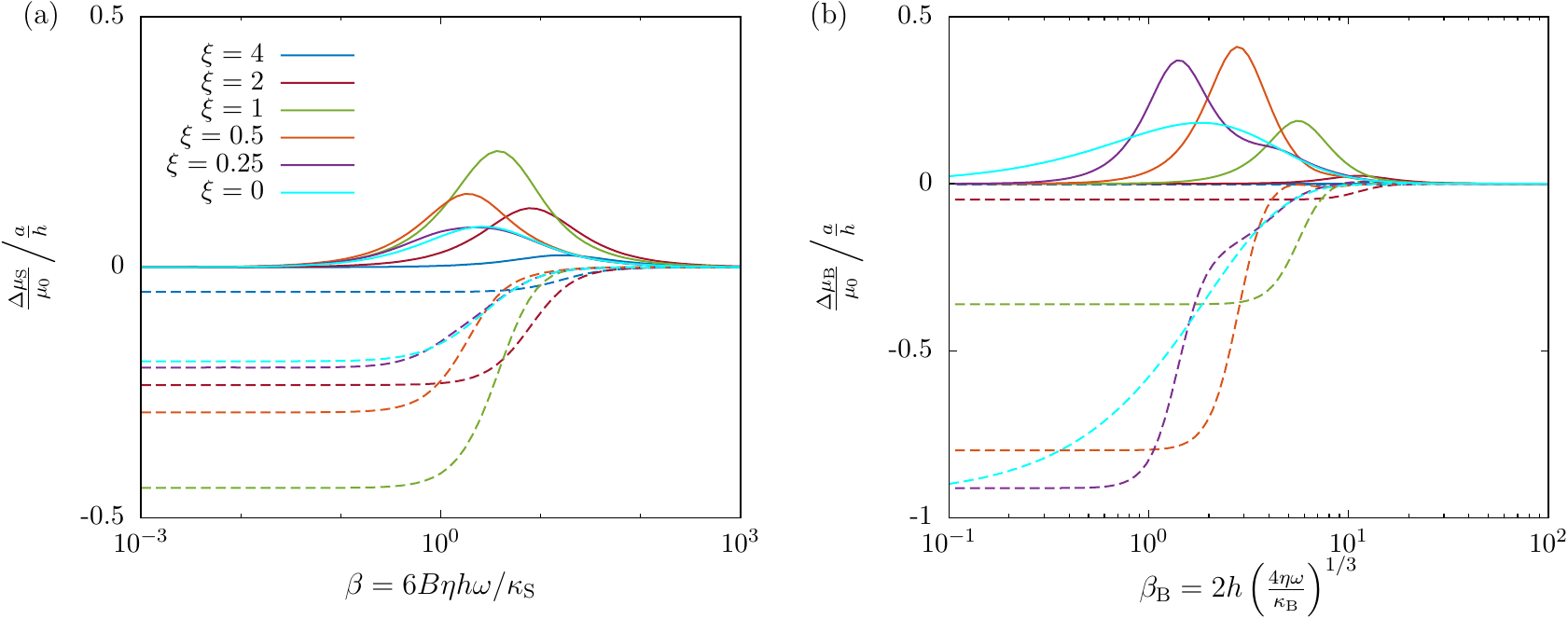}
\end{center}
\caption{(Color online) Correction factor in the hydrodynamic mobility function versus scaled frequencies for the axisymmetric motion near a finite-sized elastic membrane with (a) pure shear and (b) pure bending, for various values of $\xi=h/R$.
The curves are obtained from the semi-analytical formulas given by Eqs.~\eqref{deltaMu_Shear} and \eqref{deltaMu_Bending}, for the shear- and bending-related contributions, respectively, by taking $N=10$ in the series expansions.
Dashed and solid lines correspond to the real and imaginary parts, respectively.
The curves corresponding to~$\xi=0$ represent the mobility corrections near an infinitely extended elastic membrane given by Eqs.~\eqref{Paper_PRE16_eqn}.
}
\label{MobiCorr_ShearBending}
\end{figure}

In order to investigate the system behavior in the intermediate frequencies, we define a dimensionless number associated with shear as $\beta=2h/\alpha$, in addition to a dimensionless number associated with bending as $\betaB =2h/\alphaB$.
Accordingly, $\beta$ and $\betaB^3$ can be viewed as dimensionless frequencies associated with shear and bending deformation modes, respectively.
Moreover, we define the reduced bending modulus as $\EB = \kB/(h^2\kS)$, a parameter that quantifies the relative importance of bending and shear effects. 
For a membrane with both shear and bending, the dimensionless numbers~$\beta$ and $\betaB$ are related via
\begin{equation}
	\betaB = 2 \left( \frac{2\beta}{3B\EB} \right)^{1/3} \, .
\end{equation}

Near an infinite-extended membrane $(\xi=0)$, the shear- and bending-related contributions to the mobility corrections for the perpendicular motion has been derived is earlier work and are given by~\cite{daddi16}
\begin{subequations}\label{Paper_PRE16_eqn}
	\begin{align} 
	 \lim_{R\to\infty} \frac{\Delta {\mu}_{\mathrm{S}} }{\mu_0} &= 
	 -\frac{9}{16} \frac{a}{h} \, e^{i\beta} \mathrm{E}_4 (i\beta) \, ,\label{Paper_PRE16_eqn:perpShear}
	 \\
	 \lim_{R\to\infty} \frac{\Delta {\mu}_{\mathrm{B}}}{\mu_0} &= 
	 \frac{3i \betaB}{8} \frac{a}{h}
	  \Bigg(
	 \left( \frac{\betaB^2}{12}+\frac{i\betaB}{6} + \frac{1}{6}\right)\lambda_+
	 + \frac{\sqrt{3}}{6} (\betaB+i)\lambda_- 
	 +  \frac{5i}{2\beta_{\mathrm{B}}}
	 + e^{-i\betaB} \mathrm{E}_1(-i\betaB ) \left( \frac{\beta^2}{12} -\frac{i\betaB}{3} -\frac{1}{3}\right)
	 \Bigg) \, ,\label{Paper_PRE16_eqn:perpBen}
	\end{align}
\end{subequations}
where
\begin{equation}
 \lambda_{\pm} =  e^{-i \overline{z_{\mathrm{B}}}} \mathrm{E}_1 \left(-i\overline{z_{\mathrm{B}}} \right) \pm  e^{-i z_{\mathrm{B}}} \mathrm{E}_1 \left(-iz_{\mathrm{B}} \right) \, , 
\end{equation}
with $z_{\mathrm{B}}=\beta_\mathrm{B} \, e^{2i\pi/3}$. 
Here, bar stands for complex conjugate, and $\mathrm{E}_n$ denotes the $n$th order exponential integral~\cite{abramowitz72}.

In Fig.~\ref{MobiCorr_ShearBending}, we present the variations of the scaled mobility corrections versus the scaled frequencies for a particle moving near an idealized membrane with (a) pure shear or (b) pure bending.
Results are shown for various values of~$\xi$, which span the most likely values to be expected for a wide range of situations. 
Here, the mobilities are obtained from the semi-analytical formulas given by Eqs.~\eqref{deltaMu_Shear} and \eqref{deltaMu_Bending}, for the shear and bending-related contributions, respectively, by taking $N=10$ in the series expansions of $\chi_\mathrm{S}$ and $\chi_\mathrm{B}$.
We observe that the real parts of the mobility corrections (dashed lines) are monotonically increasing functions of frequency, whereas the imaginary parts (solid lines) exhibit typical peak structures at intermediate frequencies.
These peaks are a clear signature of the memory effect resulting from the elastic nature of the membrane. 
As $\beta, \betaB \to \infty$, the mobility corrections amount to zero, and thus one recovers the system behavior in a bulk fluid.

\section{Conclusions}
\label{sec:conclusions}

In summary, we have presented in this work an analytical theory of the axisymmetric flow induced by a Stokeslet singularity situated along the symmetry axis of a finite-sized elastic membrane of circular shape possessing resistance toward shear and bending.
We have formulated the solution of the elastohydrodynamic problem in terms of a system of dual integral equations which has then been reduced to a set of Fredholm equations of the second kind with logarithmic kernel.
Thereupon, we have provided semi-analytical expressions of the hydrodynamic mobility function for a particle translating perpendicular to an elastic membrane of finite radius.
Due to the memory effect induced by the elastic membrane, the mobility function is a frequency-dependent complex quantity that we have expressed in term  of two dimensionless parameters~$\beta$ and $\betaB$ accounting for the shear and bending deformation modes, respectively.
In the quasi-steady limit of vanishing frequency, we have shown that the hydrodynamic mobility near a finite-sized membrane is always greater in comparison to the hard disk counterpart of equal size.

Further, we have demonstrated that the bending-related contribution to the mobility correction monotonically decreases in magnitude upon increasing the dimensionless parameter~$\xi$ representing the ratio between the particle-membrane distance and the radius of the circular membrane.
In contrast to that, the shear-related part exhibits a maximum at $\xi = 1$ before to decrease in magnitude with the ratio~$\xi$.
In particular, both shear and bending effects have been found to be of about the same magnitude for $\xi=1$.
Accordingly, the system behavior has been identified to be bending and shear dominated below and above this threshold value, respectively.
In view of recent experimental advances involving controlled manipulation of particles near interfaces using optical trapping techniques, an experimental assessment of our results might be of value to future microrheology experiments involving hydrodynamically-mediated interactions near elastic interfaces.

\begin{acknowledgments}
The authors gratefully acknowledge support from the DFG (Deutsche Forschungsgemeinschaft) through the projects DA~2107/1-1 and LO~418/16-3.
\end{acknowledgments}

\appendix

\section{Derivation details of the integral equations}
\label{appendix:derivationDetails}

In this Appendix, we provide technical details regarding the derivation of the final expression of the integral equation for the shear-related part, given in the main body of the paper by Eq.~\eqref{fredholmEqn}.

By multiplying both members of Eq.~\eqref{integralEqShearTmp} by $r\, \Intd r/(s^2-r^2)^{1/2}$ and integrating with respect to the variable~$r$ in the domain~$[0,s]$, the resulting integral equation reads
\begin{equation}
 \int_0^s \frac{r\, \Intd r}{(s^2-r^2)^{1/2}} \int_r^R \frac{\chi_\mathrm{S} (t) \, \Intd t}{(t^2-r^2)^{1/2}} 
 -i\alpha \int_0^s \frac{r\, \Intd r}{(s^2-r^2)^{1/2}}  \int_0^r \frac{\chi_\mathrm{S} '(t) \, \Intd t}{(r^2-t^2)^{1/2}}  
 - \frac{i\pi\alpha}{2} \, \chi_\mathrm{S}(0) 
  = \int_0^s \frac{r w(r)\, \Intd r}{(s^2-r^2)^{1/2}}  . \label{longEq}
\end{equation}

The evaluation of the first term in the latter equation is challenging on account of the complex integration domain.
To overcome this difficulty, we employ Fubini's theorem to rewrite the double integral in a different form.
For a given integrable function~$\varphi(r,t)$ that is defined in the domain $0 \le r \le s$ and $r \le t \le R$, the theorem states that~\cite{tricomi85}
\begin{equation}
	\int_0^s \Intd r \int_r^R \varphi (r,t) \, \Intd t = 
	\int_0^s \Intd t \int_0^t \varphi (r,t) \, \Intd r +
	\int_s^R \Intd t \int_0^s \varphi (r,t) \, \Intd r \, .
\end{equation}

Accordingly, a double integration over a prismatic domain is transformed into a sum of rectangular and triangular domains that can more easily be evaluated.
Hence, after some algebra, we obtain
\begin{equation}\label{firstTermOnLHS}
 \begin{split}
  \int_0^s \frac{r\, \Intd r}{(s^2-r^2)^{1/2}} \int_r^R \frac{\chi_\mathrm{S} (t) \, \Intd t}{(t^2-r^2)^{1/2}} = \frac{1}{2} \int_0^R \chi_\mathrm{S}(t) \ln \left| \frac{s+t}{s-t} \right| \, \Intd t \, .
 \end{split}
\end{equation}

The second term in Eq.~\eqref{longEq} simplifies to
\begin{equation}
 \int_0^s \frac{r\, \Intd r}{(s^2-r^2)^{1/2}} \int_0^r \frac{\chi_\mathrm{S} '(t) \, \Intd t}{(r^2-t^2)^{1/2}} =  
  \frac{\pi}{2} \big( \chi_\mathrm{S} (s) - \chi_\mathrm{S}(0) \big)  \, ,
 \label{lastTermOnRHS}
\end{equation}
where we have used once more Fubini's theorem which, for an integrable function~$\varphi(r,t)$ that is defined in the domain $0\le r\le s$ and $0\le t\le r$, states that
\begin{equation}
	\int_0^s \Intd r \int_0^r \varphi (r,t) \, \Intd t 
	= \int_0^s \Intd t \int_t^s \varphi(r,t) \, \Intd r \, .
\end{equation}

The right-hand side can readily be evaluated as
\begin{equation}
 \int_0^s \frac{r w(r)\, \Intd r}{(s^2-r^2)^{1/2}}  = -\frac{6Fi\alpha h^2}{\pi\eta} \frac{s(h^2-s^2)}{(h^2+s^2)^4} \, . \label{TermOnRHS}
\end{equation}

\section{Recovery of the solution of the mixed boundary problem as $R\to\infty$}
\label{appendix:recoveryOFSolutionAsRgoesToInfty}

In this appendix, we show how the solution of the elastohydrodynamic problem for an infinitely-extended membrane, given by Eqs.~\eqref{f_solutionRInf}, can be recovered from the resulting integral equations of the mixed boundary value problem, in the limit when~$R\to\infty$.
For that purpose, we proceed by considering the shear and bending deformation modes independently.

\subsection{Shear contribution}

We begin with the shear related-part and demonstrate how the solution for an infinite membrane can be recovered as $R\to\infty$.
In this limit, the expressions of~$\chi_\mathrm{S}(t)$ can be obtained from Eq.~\eqref{functionF} by inverse Fourier sine transform as
\begin{equation}
 \chi_\mathrm{S}(t) = \frac{2}{\pi} \int_0^\infty q f_\mathrm{S} (q) \sin (qt) \, \Intd q \, .
\end{equation}

By substituting the latter expression into the Fredholm integral equation given by Eq.~\eqref{fredholmEqn}, and letting~$R$ goes to infinity, we obtain
\begin{equation}
  \int_0^\infty q f_\mathrm{S} (q) \sin (qs) \, \Intd q =
 \frac{Fh}{4\pi\eta} \int_0^\infty  q^3 e^{-qh} \sin (qs) \, \Intd q 
 +  \frac{1}{i\pi\alpha} \int_0^\infty \Intd t \, \ln \left| \frac{s+t}{s-t} \right|  \int_0^\infty q f_\mathrm{S} (q)\sin(qt) \, \Intd q \, ,
\end{equation}
wherein the (known) first term appearing on the right-hand side has been expressed using a Fourier representation
Using the change of variable $\lambda = (s+t)/(s-t)$ for the integrand of the second term in the latter equation, and performing the integration with respect to the variable~$t$, the integral equation can be presented in a simplified form as 
\begin{equation}
 \int_0^\infty q f_\mathrm{S} (q) \sin (qs) \, \Intd q = 
   \int_0^\infty \left( \frac{Fh}{4\pi\eta} \, q^3 e^{-qh} + \frac{f_\mathrm{S} (q)}{i\alpha}  \right)  \sin (qs) \, \Intd q \, .
\end{equation}

The solution for $f_\mathrm{S} (q)$ can be obtained by equating the Fourier components on both sides of the resulting equation.
The result after rearrangement is found to be identical to that derived using Hankel transformation given by Eq.~\eqref{f_S_solutionRInf} of the main text.

\subsection{Bending contribution}

Considering next the bending-related contribution in the limit when $R\to\infty$, the solution for~$\chi_\mathrm{B}(t)$ can be obtained by inverse Fourier since transform of Eq.~\eqref{lambdaBending} as
\begin{equation}
	\chi_\mathrm{B}(t) = \frac{2}{\pi} \int_0^\infty f_\mathrm{B}(q) \sin(qt) \, \Intd q \, .
\end{equation}

Inserting the latter equation into Eq.~\eqref{fredholmEqn_Bending}, and taking an infinite membrane radius, gives after rearranging terms, 
\begin{equation}
\int_0^\infty q^2 f_\mathrm{B}(q) \sin(qs) \, \Intd q = 
 \int_0^\infty  \left( \frac{F}{4\pi\eta} \, q^3(1+qh) e^{-qh} + \frac{1}{i\alphaB^3} \frac{f_\mathrm{B}(q)}{q} \right) \sin(qs)\, \Intd q \, ,
\end{equation}
where $b_2 = b_4 = 0$ since we have required that $\chi_\mathrm{B}(s=0) = 0$.
The solution for $f_\mathrm{S} (q)$ follows forthwith after equating the Fourier components on both sides of the latter equation.
The result is found to be identical to that derived using Hankel transformation given by Eq.~\eqref{f_B_solutionRInf} of the main body of the paper.

\section{The shear and normal stresses at the membrane}
\label{appendix:NEU}

Basing on the exact analytical expressions obtained in Sec.~\ref{sec:hydrodynamicMobility} for~$f_\mathrm{S}(q)$ and~$f_\mathrm{B} (q)$ in the limit of infinite shear and bending moduli, we provide in the following expressions for the traction jumps induced by shear and bending deformation modes in this limit.

\subsection{Shear contribution}

From the general form solution given by Eq.~\eqref{fullSolutionIntegralForm}, it follows that the tangential traction jump due to shear can be written as
\begin{equation}
	\left[ \sigma_{rz} \right] = -2\eta \int_0^\infty f_\mathrm{S} (q) J_1(qr) \, \Intd q \, .
\end{equation}

Inserting into the latter equation the expressions of~$f_\mathrm{S}(q)$ given by Eq.~\eqref{f_S}, the shear stress in the limit when~$\alpha\to\infty$ reads
\begin{equation}
	 \begin{split}
			\lim_{\alpha\to\infty} \left[ \sigma_{rz} \right] &= 
			-\frac{Fh^2}{\pi^2} \Bigg( 
			\frac{3r}{\left(h^2+r^2\right)^{5/2}} \, \arctan \left( \left( \frac{R^2-r^2}{h^2+r^2} \right)^{1/2} \right) \\
			&\qquad+\frac{8R^3\left(h^2+r^2\right)^2 
			+ \left( 2(h^2-3R^2)r^4 + (R^2-h^2)(3R^2-5h^2) r^2 - 8R^2h^4\right) \left(R^2-r^2\right)^{1/2}}{\left(h^2+r^2\right)^2 \left(R^2+h^2\right)^3 r}
			\Bigg) \, .
	\end{split} \label{sigmaRZ}
\end{equation}
Here, we have made use of the relation
\begin{equation}
	\int_{0}^{\infty} \frac{\sin (qt)}{q} \, J_1(qr) \, \Intd q = 
	\frac{1}{r} \left( t - \left(t^2-r^2\right)^{1/2} \, H(t-r) \right) \, .
\end{equation}

Consequently, the tangential traction jump for infinite shear vanishes for $r=0$ (as required by the system axisymmetry), reaches a local minimum at some intermediate distance before to decrease in magnitude as~$r \to R$. 
Accordingly, the minimum value corresponds to the maximum traction jump (in magnitude).

We denote by~${\tau_\mathrm{S}}$ the magnitude of the extremum in the membrane shear stress, defined as
\begin{equation}
	\tau_\mathrm{S} (h,R) = \left.  \lim_{\alpha\to\infty} -[\sigma_{rz}] \right|_{r=r_0} \, , 
	\quad \text{such that} \quad \left. 
	 \frac{\partial}{\partial r} \, 
	  \lim_{\alpha\to\infty}-[\sigma_{rz}]\right|_{r=r_0} = 0 \, . 
\end{equation}

Using standard optimization algorithms, such as gradient descent methods~\cite{nocedal06}, it can readily be shown that the optimum value of~$\tau_\mathrm{S}$ occurs when~$h=R$.
Accordingly, the correction to the hydrodynamic mobility is found to be larger when~$h=R$ since the shear stresses at the membrane reach an extremum value and thus causing an enhanced resistance to the motion of the particle.

In this context, the total tangential force exerted on the membrane due to shear follows readily from surface integration~\cite{morrison13}.
We obtain
\begin{equation}
	F_r =  \int_0^R [\sigma_{rz}] \, 2\pi r \,  \Intd r 
	= \frac{F}{\left(1+\xi^2\right)^{3}} \left(\left(3-\frac{16}{\pi}\right)\xi^2-1\right) \, .
\end{equation}

\subsection{Bending contribution}

The normal traction jump due to bending can be obtained by making use of Eqs.~\eqref{fullSolutionIntegralForm} as
\begin{equation}
	\left[ \sigma_{zz} \right] = 2\eta
	\int_0^\infty f_\mathrm{B} (q) J_0(qr) \, \Intd q \, , 
\end{equation}
which, upon insertion of the expression of~$f_\mathrm{B}(q)$ from Eq.~\eqref{f_B} in the limit when~$\alphaB\to\infty$ becomes
\begin{equation}
	\begin{split}
		\lim_{\alphaB \to\infty} \left[ \sigma_{zz} \right] &= 
			\frac{Fh^3}{\pi^2} \Bigg( 
			\frac{3}{\left(r^2+h^2\right)^{5/2}} \, \arctan \left( \left( \frac{R^2-r^2}{h^2+r^2} \right)^{1/2} \right) \\
			&\qquad\qquad+ \frac{\left(3h^2+R^2+2r^2\right)\left(3R^2-h^2-4r^2\right)\left(R^2-r^2\right)^{1/2}}{\left(r^2+h^2\right)^2 \left(R^2+h^2\right)^3} \Bigg) \, .
	\end{split}\label{sigmaZZ}
\end{equation}

Unlike the tangential traction jump, the normal traction jump shows a peak value at $r=0$, and decays monotonically as~$r$ increases. 

We define in an analogous way as done for shear the maximum normal stress due to bending as
\begin{equation}
	\tau_\mathrm{B} (h,R) = \left. \lim_{\alphaB\to\infty} [\sigma_{zz}] \right|_{r=0} \, . 
\end{equation}
The latter increases monotonically as the ratio~$R/h$ increases.
As a result, the bending-induced correction to the particle mobility shows a monotonic behavior and increases as the membrane size becomes larger.

For~$h < R$, the system behavior is bending dominated since $\tau_\mathrm{B} > \tau_\mathrm{S}$, while for $h > R$ the system is shear dominated since $\tau_\mathrm{B} < \tau_\mathrm{S}$.
A crossover between these two regimes occur at~$R=h$ where ~$\tau_\mathrm{S} \simeq \tau_\mathrm{B}$.

Similarly, the total normal force exerted on the membrane due to bending can be obtained by surface integration as
\begin{equation}
	F_z =  \int_0^R [\sigma_{zz}] \, 2\pi r \,  \Intd r 
	= \frac{2F}{3\pi} \left( 3\arctan \left(\frac{1}{\xi}\right) 
	+\frac{\xi \left(3+\xi^2\right) \left(1-3\xi^2\right)}{\left(1+\xi^2\right)^3}
	\right) \, .
\end{equation}

Particularly, in the limit of infinite radius of the membrane, Eqs.~\eqref{sigmaRZ} and \eqref{sigmaZZ} take the form
\begin{align}
	\lim_{\alpha, R\to\infty} \left[ \sigma_{rz} \right] &=
	-\frac{3F}{2\pi} \frac{r h^2}{\left( h^2+r^2 \right)^{5/2}} \, , \\
	\lim_{\alphaB, R\to\infty} \left[ \sigma_{zz} \right] &=
	\frac{3F}{2\pi} \frac{h^3}{\left( h^2+r^2 \right)^{5/2}} \, ,
\end{align}
both of which are in agreement with the well-known results calculated by Lorentz for an infinitely-extended planar hard wall~\cite{lorentz07, blake71}.
For this limit, the tangential and normal forces exerted at the membrane are obtained as~$F_r = -F$ and $F_z = F$.

\input{main.bbl}

\end{document}

%% file: commands.tex
\newcommand{\vect}[1]{\boldsymbol{#1}}

\newcommand{\bNabla}{\boldsymbol{\nabla}}

\newcommand{\R}{\vect{r}}

\newcommand{\RS}{\vect{r}_{\mathrm{S}}}

\newcommand{\kS}{\kappa_\mathrm{S}}
\newcommand{\kA}{\kappa_\mathrm{A}}
\newcommand{\kB}{\kappa_\mathrm{B}}

\newcommand{\EB}{E_\mathrm{B}}

\newcommand{\alphaB}{\alpha_\mathrm{B}}

\newcommand{\Intd}{\mathrm{d}}

\newcommand{\betaB}{\beta_{\mathrm{B}}}

\providecommand*{\ez}{\vect{\hat{e}}_{z}}
\providecommand*{\ex}{\vect{\hat{e}}_{x}}
\providecommand*{\ey}{\vect{\hat{e}}_{y}}

\newcommand{\vStokcom}{v^\mathrm{S}}
\newcommand{\pStok}{p^\mathrm{S}}

\newcommand{\vImcom}{v^*}
\newcommand{\pIm}{p^*}

\newcommand{\bigO}{\mathcal{O}}

%% file: main.bbl
%